\documentclass[aps,prl,reprint]{revtex4-1}
\usepackage{lipsum}
\usepackage{graphicx}
\usepackage{epstopdf}
\usepackage{dcolumn}
\usepackage{bm}
\usepackage{physics}
\usepackage{xcolor}
\usepackage{amssymb}
\usepackage{amsmath}

\begin{document}
\preprint{APS/123-QED}

\title{Quantum simulation of extended electron-phonon coupling models in a hybrid Rydberg atom setup}
\author{Jo{\~{a}}o P. Mendon{\c{c}}a}
\email{jpedromend@gmail.com}
\affiliation{%
		Faculty of Physics, University of Warsaw, Pasteura 5, 02-093 Warsaw, Poland 
	}
\author{Krzysztof Jachymski}
\affiliation{%
		Faculty of Physics, University of Warsaw, Pasteura 5, 02-093 Warsaw, Poland 
	}
\date{\today}

\begin{abstract}
    State-of-the-art experiments using Rydberg atoms can now operate with large numbers of trapped particles with tunable geometry and long coherence time. We propose a way to utilize this in a hybrid setup involving neutral ground state atoms to efficiently simulate condensed matter models featuring electron-phonon coupling. Such implementation should allow for controlling the coupling strength and range as well as the band structure of both the phonons and atoms, paving the way towards studying both static and dynamic properties of extended Hubbard-Holstein models.
    
\end{abstract}
\maketitle

{\it Introduction}. The quest for theoretical understanding of strongly correlated many-body systems is known to be extremely challenging. Classical simulation methods can become inefficient as the Hilbert space becomes too large to effectively sample. The number of parameters needed to describe and store a quantum state grows exponentially with the system size. Quantum computers hold the promise to overcome this difficulty, but near-term devices cannot be expected to provide the needed number of logical qubits and sufficient circuit depth. For these reasons, analog quantum simulators are among the most promising tools to study ground state properties as well as dynamics of interacting quantum systems~\cite{Cirac2012,Georgescu2014,Fraxanet2022}. As the approach of analog simulation is to construct a controllable system that can reproduce the physics of a different one, it is nonuniversal and thus the details of experimental implementation matter. The platforms for quantum simulation must feature some versatility in tuning the system parameters, scalability in the number of qubits, and reliable measurement schemes. Over the past two decades, a wide range of promising quantum devices that may have some of the desired properties has emerged~\cite{QS_PRX2021}.
Among the various platforms available, ultracold neutral atoms~\cite{Bloch2012} and Rydberg atom arrays ~\cite{Bernien2017} received a lot of attention in this respect. A hybrid approach involving a combination of setups can be promising as well, allowing for easier implementation of more complex systems with potentially independently tunable properties~\cite{Bissbort2013}.

Strongly correlated materials can feature a competition between electron-electron and electron-phonon interactions which drive the system towards different ordered phases. Phonon-mediated attraction between electrons can enhance fermion pairing even when the Coulomb repulsion is strong~\cite{Alexandrov1981,lanzara2001evidence}. In such materials, as well as high-$T_c$ superconductors, the Hamiltonians cannot generally be treated by perturbative methods because of the lack of a small parameter. Computational methods such as exact diagonalization (ED), quantum Monte Carlo (QMC), and density matrix renormalization group (DMRG), greatly advanced the understanding of many-body systems. However, as the phonon Hilbert space has to be truncated, numerical studies typically allow only a small phonon number and system size.

A number of theoretical proposals for quantum simulation of electron-phonon models using molecules as well as ions exists~\cite{Pupillo2008,Ortner2009,Hague2012,Jachymski2020}. However, they have stringent requirements and can lack versatility. For example, in order to crystallize the molecules one needs extremely low temperatures, while in ion-atom systems the relevant energy scales are quite separated. Here we focus on a different type of mixture involving an array of Rydberg atoms and a ground state gas.
In most experimental realizations, the Rydberg states are repelled by optical traps and the laser field must be turned off during experiments. However, recent developments allow for keeping the tweezer array on as well as to achieve state-insensitive traps~\cite{Wilson2022,Kuzmich2022}, leading to long lifetimes and opening the door towards a new simulation platform.

\begin{figure*}[ht]
    \centering
    \includegraphics[width=0.9\linewidth]{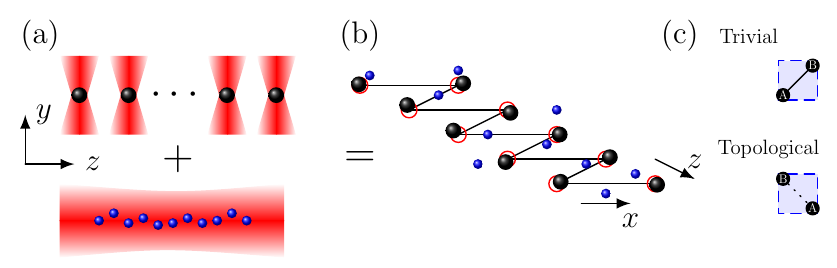}
    \caption{(a) Top: Red-detuned optical tweezers generating the Rydberg lattice; bottom: state-insensitive trap for the ground state atomic cloud. (b) The proposed platform. Strongly interacting Rydberg atoms (black balls) are trapped in an array of tweezers (red circles). The cloud neutral atoms (blue balls) is placed in a periodic lattice potential due to the Rydberg chain. The lines are to guide the eyes. (c) The unit cells generating two lattice geometries with the base atoms labeled as $A$ and $B$; the interactions shown here are strong (weak) for the trivial (topological) scenario, respectively.}
    \label{fig:1}
\end{figure*}

In this work, we extend this notion and provide a scheme for quantum simulation of strongly correlated many-body systems. To begin, we study the phonon spectrum of a Rydberg chain showcasing its tunability. Then we argue that the array can be seen as a periodic potential for the neutral atoms (see Fig.~\ref{fig:1}(a)). Following that, we derive and study the full system Hamiltonian which contains atom-phonon coupling. Finally, we discuss further prospects for quantum simulation in this setup.

The simplest and widely used Hamiltonian that takes into account both electron-electron and electron-phonon interactions is the Hubbard-Holstein (HH) model,
\begin{align}\label{eq:HH}
    H &= -t\sum_{\langle i,j\rangle\sigma}(c^{\dagger}_{i\sigma}c_{j\sigma} + \textrm{H.c.}) + U\sum_{i}n_{i\uparrow}n_{i\downarrow} 
    \nonumber\\ &+ \omega_0\sum_i b_i^{\dagger}b_i + g\sum_{i\sigma}n_{i\sigma}(b_i^{\dagger}+b_i)\, ,
\end{align}
where operators $c_{i\sigma}$ annihilate an electron with spin $\sigma$ at site $i$ with $n_{i\sigma}=c^\dagger_{i\sigma} c_{i\sigma}$ while $b_i$ govern the local phonons with a single frequency $\omega_0$, and $g$ is the onsite coupling strength. The model features surprisingly rich physics as phonon-induced interactions compete with the Hubbard term~\cite{Lang1963}, leading to emergence of two insulating orders with a metallic phase appearing at their interplay~\cite{Fehske1996,Takada2003,Clay2005,Hohenadler2007,Campbell2014,Yin2015,Wang2020}. 
The model can be easily extended which considerably increases its complexity. For instance, one could modify the last term of Eq.~\eqref{eq:HH} by introducing nonlocal electron-phonon coupling $\mathcal{H}_{e-ph}$. Within the second quantization, the extended version of the model is written as
\begin{equation}
    \mathcal{H}_{e-ph} = \sum_{\vb{q}i} g_{\vb{q}i} (b_{\vb{q}}^{\dagger}+b_{\vb{q}})\rho_{i},
\end{equation}
where $\rho_i$ is the electron density and $b_{\vb{q}}$ is the reciprocal representation of the local phonon $b_i$. While retaining a simple form, this term is rather general and can describe long-range couplings with nontrivial structure.

Furthermore, the phonon as well as electron dispersion can be replaced by a richer and more realistic structure. To showcase this, here we use a zig-zag configuration of Rydberg atoms with anisotropic interactions instead of a more standard cubic arrangement. This choice is motivated by a recent experiment~\cite{Leseleuc2019} which emulated the physics of the Su-Schrieffer-Heger model with topological edge states in a similar setup.

\textit{Topological Rydberg lattice.} 
In our approach the Rydberg atoms are constantly individually trapped by an array of harmonic potentials (optical tweezers) and interact with each other via dipolar interactions which can be induced with external electric field. As shown in Fig.~\ref{fig:1}(b), the traps are located at fixed positions forming a zig-zag chain in the $x$--$z$ plane. The atom positions are given by $\mathbf{R}_{n\alpha}(t)=\mathbf{R}_n + \boldsymbol{\rho}_{\alpha} + \mathbf{u}_{n\alpha}(t)$, where $\mathbf{R}_n=na\hat{\mathbf{z}}$ are the unit cell positions, $\vb*{\rho}_\alpha$ labels the atoms within a cell with $\alpha=A,B$ being the base atom label and $n$ the cell position index, and $\mathbf{u}_{n\alpha}(t)$ the time-dependent displacements.
The potential energy is written as
\begin{align}
    V &= \frac{1}{2}\sum_{n,\alpha}M_{\alpha}[\vb*{\nu}_n(\vb{R}_{n\alpha}-\Bar{\vb{R}}_{n\alpha})]^2 \nonumber\\
    &+\sum_{\substack{nm\alpha\beta\\(n,\alpha)\neq(m,\beta)}} \frac{V_{\textrm{dd}}}{|\vb{R}_{n\alpha,m\beta}|^3}\left[1 - 3(\vu{m}\cdot \vu{R}_{n\alpha,m\beta})^2\right],
\end{align}
where $\vu{R}_{n\alpha,m\beta}=\vu{R}_{n\alpha}-\vu{R}_{m\beta}$. In a finite zig-zag chain with anisotropic couplings between the sites, there are two distinct geometries encoded in $\vb*{\rho}_{\alpha}$~\cite{Leseleuc2019,Sompet2022} corresponding to forming pairs of dimers and leaving out two unpaired atoms at the chain edges. For the trivial configuration, we define $\boldsymbol{\rho}_{A,B}= \mp \Delta/2\hat{\mathbf{x}} \mp d/2\hat{\mathbf{z}}$, and $\boldsymbol{\rho}_{A,B}= \mp \Delta/2\hat{\mathbf{x}} \pm d/2\hat{\mathbf{z}}$ for the topological one (see Fig.~\ref{fig:1}(c)). For simplicity, we fix $\vb*{\nu}_n=\vb*{\nu}$ with $\nu_{x,y,z}=\nu$ and choose $M_{\alpha}=M$. Current stat-of-the-art experiments enable an arbitrary 3D setting of the tweezer traps with separations of the order of a few $\mu$m. The trap frequencies can be varied as well  and are typically in the~$\sim$kHz regime.
The interactions between the atoms depends on the choice of Rydberg states and can remain strong over the typical trap separation. Furthermore, it can be precisely tuned by using external electromagnetic fields, for instance by inducing and orienting the dipole moments~\cite{Weber2017}. If we choose $\vu{m}=(0,1,0)$ (out of the plane), the interactions become isotropic. Following \cite{Leseleuc2019}, let us instead fix $\theta=\theta_m=\cos[-1](1/\sqrt{3})$ where the interaction along the same sublattice vanishes as $\vu{m}\cdot \vu{R}_{n\alpha,m\beta}=\vu{z}$ and $\theta=\theta_m$. 

We now turn to the phonon structure of the Rydberg chain, expanding the potential energy to second order around the equilibrium configuration. The effective classical Hamiltonian is given by
\begin{equation}
    H_{\textrm{eff}} = \sum_{n=1}^{N_c}\sum_{\alpha=1}^{N_b}\sum_{i=x,y,z} \frac{P_{n\alpha i}^2}{2M_{\alpha}} + \frac{1}{2}\sum_{nm}\sum_{\alpha\beta}\sum_{ij} u_{n\alpha i}D_{n\alpha i}^{m\beta j}u_{m\beta j},
\end{equation}
with $N_c$ being the number of cells and $N_b=2$ the number of base atoms. The harmonic matrix $D$ is given by the second derivatives of the potential at equilibrium.
Interestingly, there is only one relevant interaction length scale $\ell^5 = 3V_{dd}/M\omega^2$ within this approximation. It is widely tunable by means of the trap frequencies as well as the choice of the Rydberg level. We further express all other length scales describing the array geometry in units of $\ell$ and energies in the corresponding characteristic units $M\omega^2 l^2$. 
One can now obtain the phonon spectrum by diagonalizing the $3N_cN_b\times 3N_cN_b$ harmonic matrix. On the other hand, if the translational invariance is applicable, $D_{\alpha i}^{\beta j}(|\vb{R}_n-\vb{R}_m|)=D_{\alpha i}^{\beta j}(\vb{R}_p)$ and one can reduce the problem in the quasimomemtum space to the $3N_b\times 3N_b$ dynamical matrix~\cite{mahan2013book} $\Tilde{D}_{\alpha i}^{\beta j}(\vb{q}) = \sum_n D_{\alpha i}^{\beta j}(\vb{R}_n)e^{i\vb{q}\cdot\vb{R}_n}$.
Due to the chosen geometry, in our example the quasimomentum is $\vb{q}=q\vu{z}$ as the system is quasi-1D.

\begin{figure}[t]
    \centering
    \includegraphics[width=\linewidth]{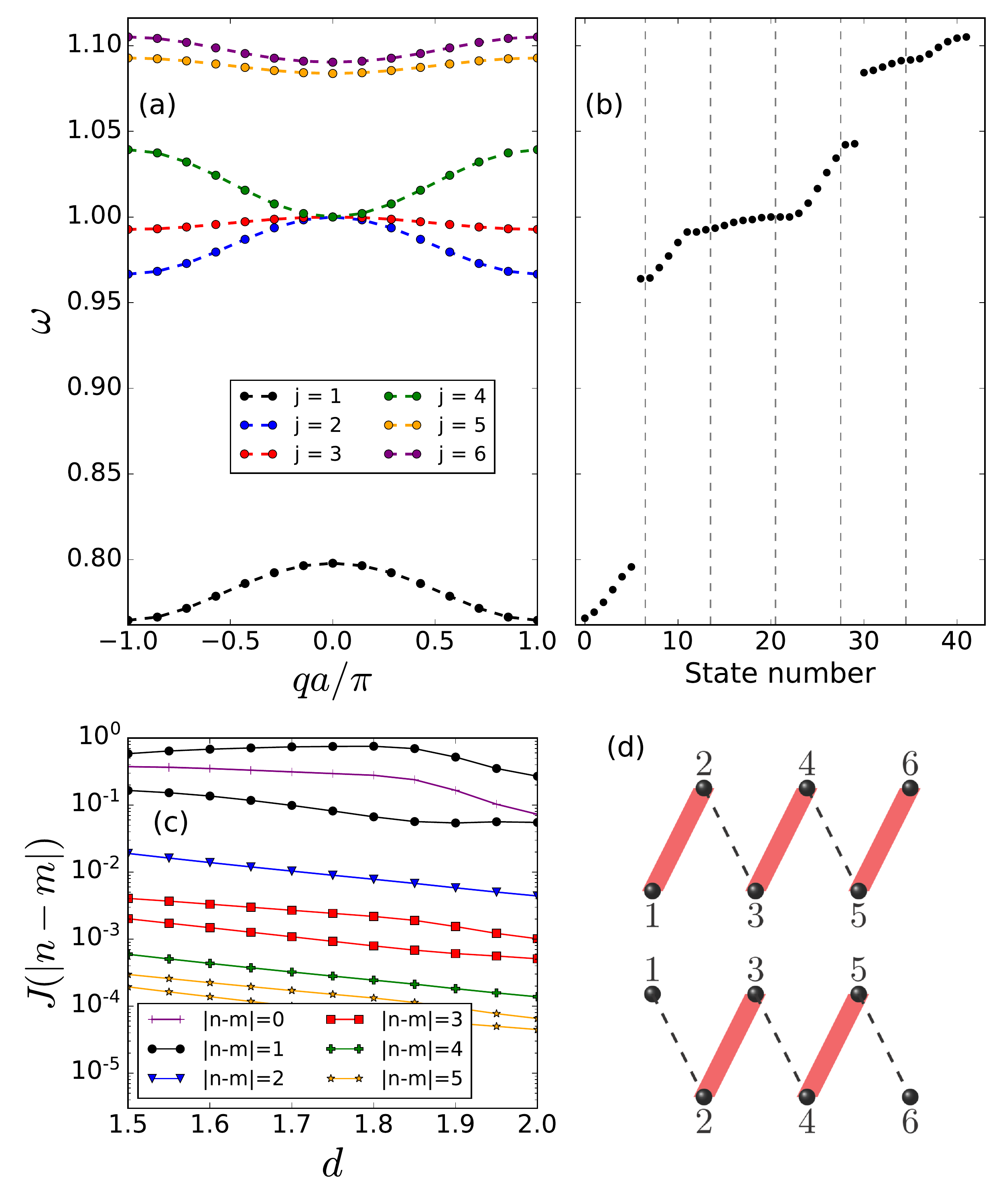}
    \caption{The phonon dispersion relation for (a) the translationally invariant case, (b) the topological band structure of the finite chain for $N=14,d=2,\Delta=1,a=2d,\theta=\theta_m,$, and $\phi=0$, (c) couplings between local phonons as a function of $d$ for different pairs $n,m$, and (d) trivial and topological geometries where the lines depict the strong (red thick line) and weak (black dashed line) links (see text).}
    \label{fig:2}
\end{figure}

In the trivial/dimerized configuration, the finite chain has translational symmetry. Figure \ref{fig:2}(a) shows its exemplary dispersion relation in the quasimomentum space. The system is widely tunable in terms of band geometry and shows a rich behavior which exhibits concavity changes and multiple band crossings~\cite{SupMat}. 
For the topological configuration, the translational invariance can no longer be assumed. In Figure \ref{fig:2}(b) we show the dispersion relation calculated from direct diagonalization of the harmonic matrix. We can see three points which are disconnected from the bands,  which are the edge states, one in the first and two in the fifth band. 
To gain more insight, following~\cite{bissbort2016} we can rewrite the phonon Hamiltonian in terms of local phonon operators
\begin{align}
    H = \frac{1}{2}\sum_{nmij}\bigg[ h_{nm}^{ij}&( b_{n,i}^{\dagger}b_{m,j} + b_{n,i}b_{m,j}^{\dagger}) \nonumber\\ + &g_{nm}^{ij}( b_{n,i}b_{m,j} + b_{n,i}^{\dagger}b_{m,j}^{\dagger}) \bigg],
\end{align}
where the displacements are quantized as $u_{n,i}=\sqrt{\frac{1}{2M\Omega_{n,i}}}(b_{n,i}+b_{n,i}^{\dagger})$, with the local frequency $\Omega_{n,i}$ defined as $D_{nn}^{ii}=M\Omega_{n,i}^2$. Furthermore, the couplings are given by the matrix $g$ with $g_{nm}^{ij} = (1-\delta_{nm}\delta_{ij})D_{nm}^{ij}/2M\sqrt{\Omega_{n,i}\Omega_{m,j}}$ and $h_{nm}^{ij} = \delta_{nm}\delta_{ij}\Omega_{n,i} + g_{nm}^{ij}$, with $n,m \in [0,N]$ now being the overall atom index where $N=N_b N_c$ (see Fig.~\ref{fig:2}(d)). The $g$ matrix has $9 N^2$ elements describing nine possible couplings between each pair.~\cite{SupMat}. For visualization we show in Fig.~\ref{fig:2}(c) the quantity $J(|n-m|)=\sum_{ij} g_{nm}^{ij}$ for distinct values of $|n-m|$ and $d$. The terms with odd $|n-m|$ describe coupling between two different legs with two cases corresponding to the majority of weak or strong bonds between the sites. Such staggered nature can be directly associated with the SSH model. Even $|n-m|$ provides the interactions along the same sublattice (see Fig.~\ref{fig:2}(d)) which does not vanish even at the magic angle $\theta_m$ due to displacement from equilibrium. As the trap separation $d$ decreases, interactions in general become stronger. However, the intracell couplings feature a maximum near $d\approx 1.8$, which is related to the system geometry and coincides with the concavity change of the lowest band, indicating the presence of geometric frustration.

We have so far demonstrated that phonons in Rydberg atom arrays are capable of simulating complex solid state systems with highly tunable band structure. Extension to a two-dimensional setup would enable the occurrence of chiral edge states. Implementation of driving into the system with lasers can induce additional nonequilibrium dynamics~\cite{Gambetta2020}.

\begin{figure*}[ht]
    \centering
    \includegraphics[width=0.4\linewidth]{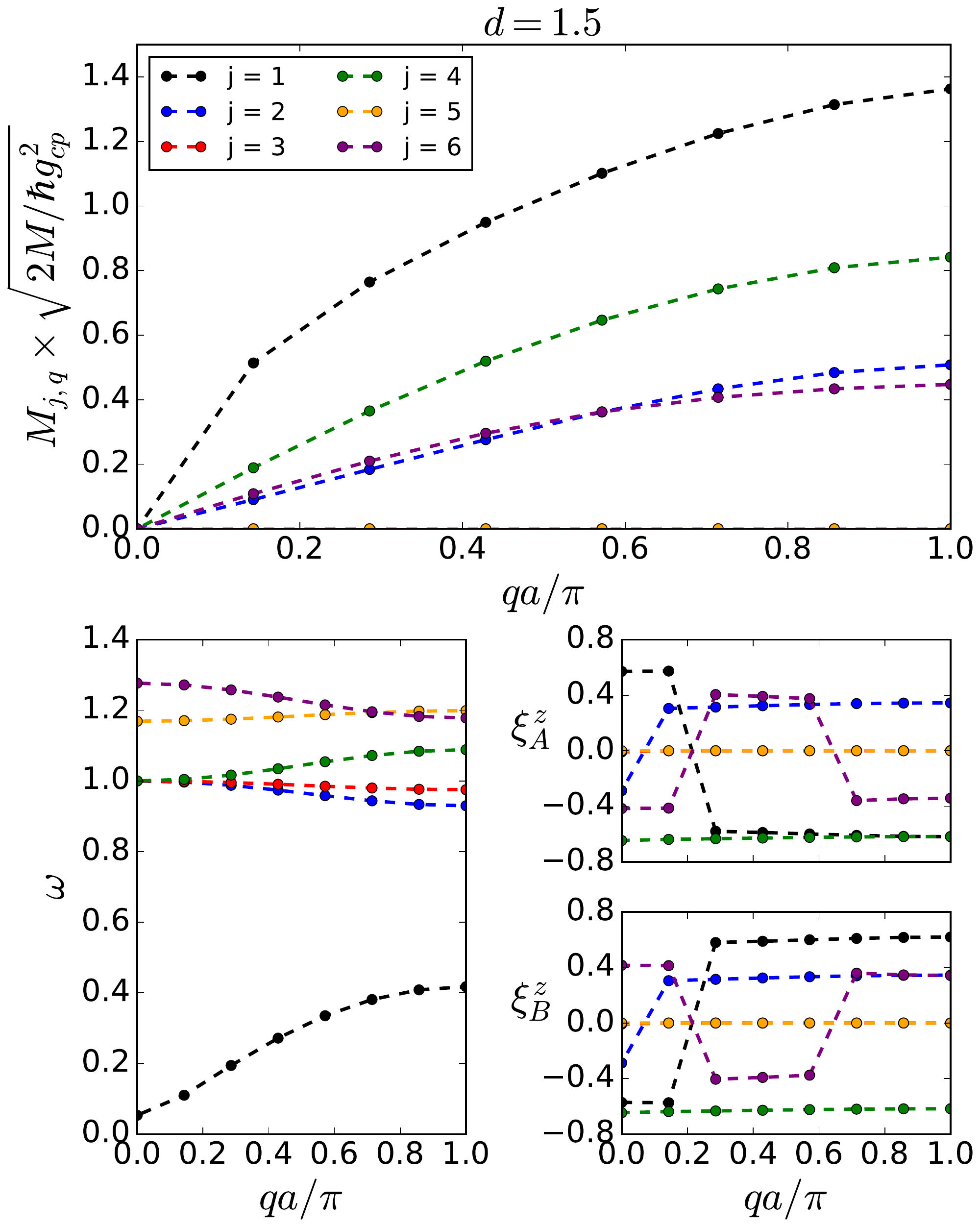}
    \includegraphics[width=0.4\linewidth]{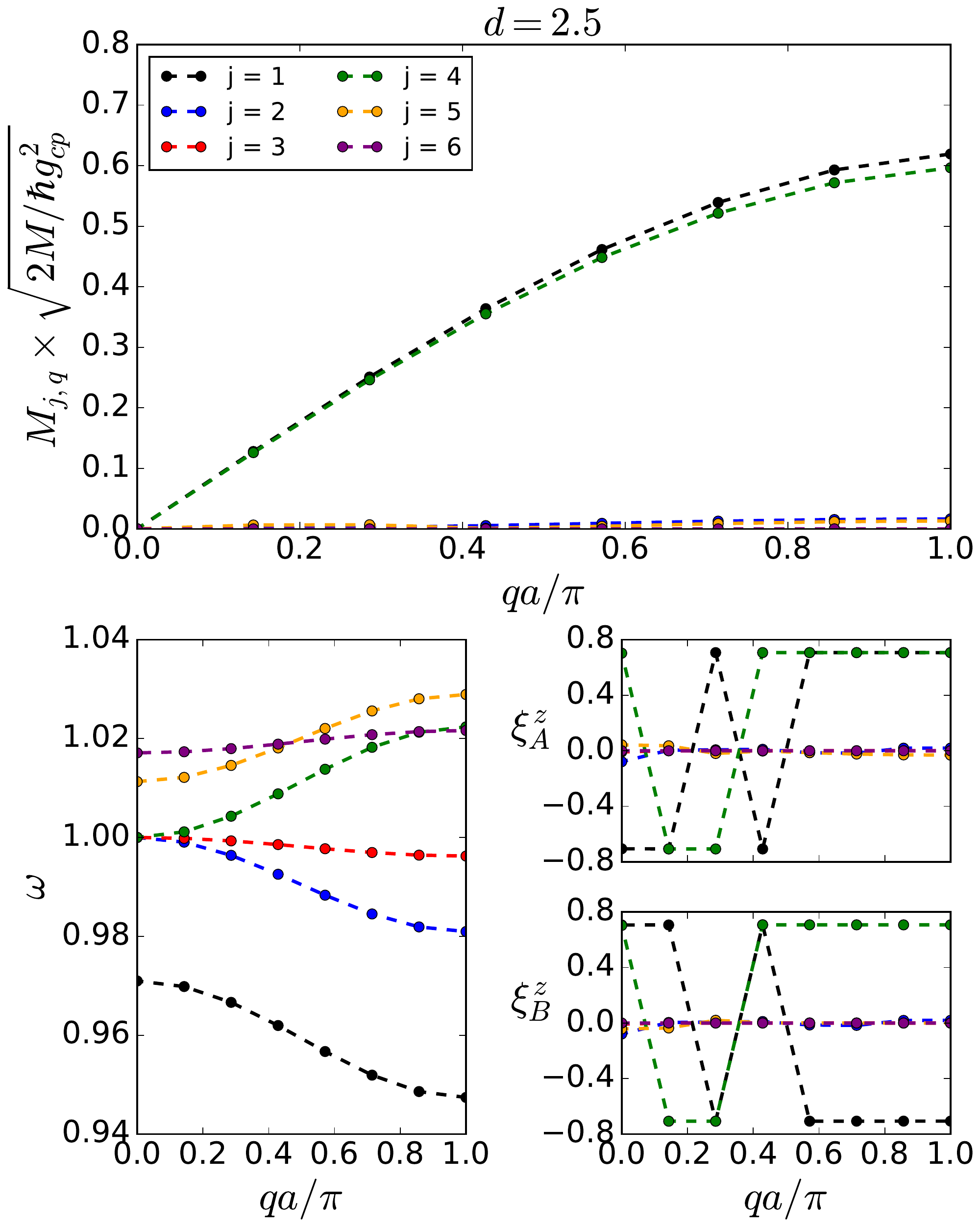}
    \caption{Interaction strength $M_{j,q}$ for $d=1.5$ and $d=2.5$. We show two examples where there is a clear distinction between multi-band and two-band regimes. The corresponding phonon spectra and eigenvector component in $z$ direction are shown in the bottom panels. Here $N=14,\Delta=1,a=2d,\theta=\theta_m,\phi=0$.}
    \label{fig:3}
\end{figure*}

\textit{Ground state atoms.} We proceed to adding the second subsystem composed of ground state atoms. Their van der Waals interaction with the Rydberg atoms can be well described by a Fermi pseudopotential with magnitude $g_{cp}$. For the whole array, $H_{\textrm{Ry-a}}=\sum_{n,m}\sum_{\alpha} V_{\textrm{Ry-a}}(\vb{r}_n-\vb{R}_{m\alpha})$. This can be expanded in power series in Rydberg displacements $u$, providing a constant term which forms the periodic potential for the atoms while the atom-phonon interaction is encoded in the first order term. Within the tight binding approximation, the bare atomic Hamiltonian without phonons is described by the Hubbard model
\begin{equation}
    \mathcal{H}_{\textrm{a}} = -t\sum_{\langle i,j\rangle\sigma}(c^{\dagger}_{i\sigma}c_{j\sigma} + \textrm{H.c.}) + U\sum_{i}n_{i\uparrow}n_{i\downarrow} .
\end{equation}
The atom-phonon Hamiltonian is obtained by the second quantization of the phonon relevant part of the interaction Hamiltonian, $H_{\textrm{Ry-a}}^{(1)}$, as shown in \cite{SupMat}.
%
For our neutral atomic system, we restrict to the lowest Bloch band and tight binding approximation.
We used a Gaussian approximation for the lattice Wannier functions, which for composite lattices may need to be modified~\cite{Marzari1997,Ganczarek2014,Negretti2014} but does not qualitatively impact our results. 
Overall, the atom-phonon Hamiltonian is then given by
\begin{equation}
    \mathcal{H}_{\textrm{a-ph}} = \sum_{\vb{q},\vb{k},j,\sigma} \frac{1}{\sqrt{N}}
    M_{j,\vb{q}} (b_{j,\vb{q}} + b_{j,\vb{-q}}^{\dagger}) c_{\vb{k+q},\sigma}^{\dagger} c_{\vb{k},\sigma},
\end{equation}
where the interaction strength is given by
\begin{equation}\label{eq:M}
    M_{j,\vb{q}} = \sum_{\alpha} \sqrt{\frac{\hslash g_{cp}^2}{2M_{\alpha}\omega_{j}(\vb{q})}} \vb{q}\cdot\vb*{\xi}_{\alpha}^{(j)}(\vb{q}) e^{-i\vb{q}\cdot\vb*{\rho}_{\alpha}} \rho_0(\vb{q}),
\end{equation}
with $\rho_0=\int \dd\vb{r} e^{i\vb{q}\cdot\vb{r}} |\phi_0(\vb{r})|^2$ and $\phi_0(\vb{r})$ being the Wannier function of the lowest Bloch band.

In a zig-zag chain where $\vb{q}=q\vu{z}$, the dot product in $M_{j\vb{q}}$ restricts the couplings and as a result transverse states without a $z$ component do not interact with the atoms. 
We observe that the interaction strength is highly dependent on the phonon energy and structure. Indeed, the other functions appearing in Eq.~\ref{eq:M} will mostly affect the overall strength the coupling profile. In order to design the desired interaction between ground state atoms and phonons one thus needs to focus on the phonon part. The proposed Rydberg lattice setup turns to be perfect for such situation since it is a rich and highly controllable platform as discussed earlier.

In Fig. \ref{fig:3} we show two exemplary situations in which the coupling term crosses over from a two-band to multi-band structure. Its overall magnitude can be easily controlled independently by manipulating the Rydberg-atom interaction strength. For a better understanding, we show in the bottom panels of Fig.~\ref{fig:3} the phonon spectrum and the $z$ component of the corresponding states. 
As we can see, when $d$ is big compared with the other lengths, only the first and 4th band are generating noticeable oscillations in $z$--direction, causing a two-band interaction strength. Decreasing the distance between the Rydberg atoms, other bands also acquire vibrational components along the $z$ axis. This is the limitation quasi-1D platform as only longitudinal phonons can contribute.

\textit{Discussion.}  As discussed above, the while system is described by an extended Hubbard-Holstein model
\begin{align}
    \mathcal{H}_{\mathrm{eff}} = &-t\sum_{\langle i,j\rangle\sigma}(c^{\dagger}_{i\sigma}c_{j\sigma} + \textrm{H.c.}) + U\sum_{i}n_{i\uparrow}n_{i\downarrow} 
    \nonumber\\
    &+ \sum_{j,\vb{q}}\hslash\omega_{j}(\vb{q}) {b}_{j,\vb{q}}^{\dagger} {b}_{j,\vb{q}}
    \nonumber \\ &+ \sum_{\vb{q},\vb{k},j,\sigma} \frac{1}{\sqrt{N}}
    M_{j,\vb{q}} (b_{j,\vb{q}} + b_{j,\vb{-q}}^{\dagger}) c_{\vb{k+q},\sigma}^{\dagger} c_{\vb{k},\sigma},
\end{align}
with nontrivial phonon structure including the possibility for topological bands and nonlocal couplings. Phonon-induced interactions can induce long-range attraction between the fermions and enhance the formation of pairs and conductivity. While one can study the limiting cases of weak or very strong coupling analytically, e.g. by means of a generalized Lang-Firsov transformation~\cite{Lang1963,Hohenadler2007,Ortner2009,Yin2015}, we stress that due to the competition of terms the problem calls for a thorough numerical study in order to verify and interpret the quantum simulator output.

Experimentally, the system provides a number of possible measurements such as time-of-flight and {\it in situ} imaging of atoms, as well as phonon state tomography of the chain. Studying the dynamics after a quench or under driving should does not impose any additional challenge.

Finally, let us mention that a class of extended Hubbard-Holstein models could also be studied with a possibly simpler system in the spirit of variational quantum simulation~\cite{Kokail2019,Meth2022}. In this scenario one would use an ansatz for the phonon part in order to calculate the effective fermionic Hamiltonian classically, and then find the ground state of the reduced system by implementing it in experiment and obtaining a new candidate phonon state for the next iteration.

\textit{Conclusions.} We have proposed a highly tunable experimental platform for simulation of compound quantum systems. It can be utilized for exploration of phase diagrams of extended Hubbard models in various geometric arrangements, possibly to study the onset of bipolaronic superconductivity. In the future, extensions to 2D and 3D structures and designing the system to exhibit flat bands and edge states seems particularly promising~\cite{Shi2013,Reshodko2019,Salamon2020,Knorzer2022,DiLiberto2022,Sompet2022}, with exciting prospects for nonequlibrium dynamics and phonon driving related to recent breakthrough results on transient superconductivity~\cite{Cavalleri2018,Babadi2017}.
 
This work was supported by the Polish National Agency for Academic Exchange (NAWA) via the Polish Returns 2019 programme.

\bibliographystyle{apsrev4-1}
\bibliography{refs.bib}{}

\begin{thebibliography}{42}%
\makeatletter
\providecommand \@ifxundefined [1]{%
 \@ifx{#1\undefined}
}%
\providecommand \@ifnum [1]{%
 \ifnum #1\expandafter \@firstoftwo
 \else \expandafter \@secondoftwo
 \fi
}%
\providecommand \@ifx [1]{%
 \ifx #1\expandafter \@firstoftwo
 \else \expandafter \@secondoftwo
 \fi
}%
\providecommand \natexlab [1]{#1}%
\providecommand \enquote  [1]{``#1''}%
\providecommand \bibnamefont  [1]{#1}%
\providecommand \bibfnamefont [1]{#1}%
\providecommand \citenamefont [1]{#1}%
\providecommand \href@noop [0]{\@secondoftwo}%
\providecommand \href [0]{\begingroup \@sanitize@url \@href}%
\providecommand \@href[1]{\@@startlink{#1}\@@href}%
\providecommand \@@href[1]{\endgroup#1\@@endlink}%
\providecommand \@sanitize@url [0]{\catcode `\\12\catcode `\$12\catcode
  `\&12\catcode `\#12\catcode `\^12\catcode `\_12\catcode `\%12\relax}%
\providecommand \@@startlink[1]{}%
\providecommand \@@endlink[0]{}%
\providecommand \url  [0]{\begingroup\@sanitize@url \@url }%
\providecommand \@url [1]{\endgroup\@href {#1}{\urlprefix }}%
\providecommand \urlprefix  [0]{URL }%
\providecommand \Eprint [0]{\href }%
\providecommand \doibase [0]{http://dx.doi.org/}%
\providecommand \selectlanguage [0]{\@gobble}%
\providecommand \bibinfo  [0]{\@secondoftwo}%
\providecommand \bibfield  [0]{\@secondoftwo}%
\providecommand \translation [1]{[#1]}%
\providecommand \BibitemOpen [0]{}%
\providecommand \bibitemStop [0]{}%
\providecommand \bibitemNoStop [0]{.\EOS\space}%
\providecommand \EOS [0]{\spacefactor3000\relax}%
\providecommand \BibitemShut  [1]{\csname bibitem#1\endcsname}%
\let\auto@bib@innerbib\@empty
\bibitem [{\citenamefont {Cirac}\ and\ \citenamefont
  {Zoller}(2012)}]{Cirac2012}%
  \BibitemOpen
  \bibfield  {author} {\bibinfo {author} {\bibfnamefont {J.~I.}\ \bibnamefont
  {Cirac}}\ and\ \bibinfo {author} {\bibfnamefont {P.}~\bibnamefont {Zoller}},\
  }\href@noop {} {\bibfield  {journal} {\bibinfo  {journal} {Nature physics}\
  }\textbf {\bibinfo {volume} {8}},\ \bibinfo {pages} {264} (\bibinfo {year}
  {2012})}\BibitemShut {NoStop}%
\bibitem [{\citenamefont {Georgescu}\ \emph {et~al.}(2014)\citenamefont
  {Georgescu}, \citenamefont {Ashhab},\ and\ \citenamefont
  {Nori}}]{Georgescu2014}%
  \BibitemOpen
  \bibfield  {author} {\bibinfo {author} {\bibfnamefont {I.~M.}\ \bibnamefont
  {Georgescu}}, \bibinfo {author} {\bibfnamefont {S.}~\bibnamefont {Ashhab}}, \
  and\ \bibinfo {author} {\bibfnamefont {F.}~\bibnamefont {Nori}},\ }\href@noop
  {} {\bibfield  {journal} {\bibinfo  {journal} {Reviews of Modern Physics}\
  }\textbf {\bibinfo {volume} {86}},\ \bibinfo {pages} {153} (\bibinfo {year}
  {2014})}\BibitemShut {NoStop}%
\bibitem [{\citenamefont {Fraxanet}\ \emph {et~al.}(2022)\citenamefont
  {Fraxanet}, \citenamefont {Salamon},\ and\ \citenamefont
  {Lewenstein}}]{Fraxanet2022}%
  \BibitemOpen
  \bibfield  {author} {\bibinfo {author} {\bibfnamefont {J.}~\bibnamefont
  {Fraxanet}}, \bibinfo {author} {\bibfnamefont {T.}~\bibnamefont {Salamon}}, \
  and\ \bibinfo {author} {\bibfnamefont {M.}~\bibnamefont {Lewenstein}},\
  }\href@noop {} {\bibfield  {journal} {\bibinfo  {journal} {arXiv preprint
  arXiv:2204.08905}\ } (\bibinfo {year} {2022})}\BibitemShut {NoStop}%
\bibitem [{\citenamefont {Altman}\ \emph {et~al.}(2021)\citenamefont {Altman},
  \citenamefont {Brown}, \citenamefont {Carleo}, \citenamefont {Carr},
  \citenamefont {Demler}, \citenamefont {Chin}, \citenamefont {DeMarco},
  \citenamefont {Economou}, \citenamefont {Eriksson}, \citenamefont {Fu},
  \citenamefont {Greiner}, \citenamefont {Hazzard}, \citenamefont {Hulet},
  \citenamefont {Koll\'ar}, \citenamefont {Lev}, \citenamefont {Lukin},
  \citenamefont {Ma}, \citenamefont {Mi}, \citenamefont {Misra}, \citenamefont
  {Monroe}, \citenamefont {Murch}, \citenamefont {Nazario}, \citenamefont {Ni},
  \citenamefont {Potter}, \citenamefont {Roushan}, \citenamefont {Saffman},
  \citenamefont {Schleier-Smith}, \citenamefont {Siddiqi}, \citenamefont
  {Simmonds}, \citenamefont {Singh}, \citenamefont {Spielman}, \citenamefont
  {Temme}, \citenamefont {Weiss}, \citenamefont {Vu\ifmmode \check{c}\else
  \v{c}\fi{}kovi\ifmmode~\acute{c}\else \'{c}\fi{}}, \citenamefont
  {Vuleti\ifmmode~\acute{c}\else \'{c}\fi{}}, \citenamefont {Ye},\ and\
  \citenamefont {Zwierlein}}]{QS_PRX2021}%
  \BibitemOpen
  \bibfield  {author} {\bibinfo {author} {\bibfnamefont {E.}~\bibnamefont
  {Altman}}, \bibinfo {author} {\bibfnamefont {K.~R.}\ \bibnamefont {Brown}},
  \bibinfo {author} {\bibfnamefont {G.}~\bibnamefont {Carleo}}, \bibinfo
  {author} {\bibfnamefont {L.~D.}\ \bibnamefont {Carr}}, \bibinfo {author}
  {\bibfnamefont {E.}~\bibnamefont {Demler}}, \bibinfo {author} {\bibfnamefont
  {C.}~\bibnamefont {Chin}}, \bibinfo {author} {\bibfnamefont {B.}~\bibnamefont
  {DeMarco}}, \bibinfo {author} {\bibfnamefont {S.~E.}\ \bibnamefont
  {Economou}}, \bibinfo {author} {\bibfnamefont {M.~A.}\ \bibnamefont
  {Eriksson}}, \bibinfo {author} {\bibfnamefont {K.-M.~C.}\ \bibnamefont {Fu}},
  \bibinfo {author} {\bibfnamefont {M.}~\bibnamefont {Greiner}}, \bibinfo
  {author} {\bibfnamefont {K.~R.}\ \bibnamefont {Hazzard}}, \bibinfo {author}
  {\bibfnamefont {R.~G.}\ \bibnamefont {Hulet}}, \bibinfo {author}
  {\bibfnamefont {A.~J.}\ \bibnamefont {Koll\'ar}}, \bibinfo {author}
  {\bibfnamefont {B.~L.}\ \bibnamefont {Lev}}, \bibinfo {author} {\bibfnamefont
  {M.~D.}\ \bibnamefont {Lukin}}, \bibinfo {author} {\bibfnamefont
  {R.}~\bibnamefont {Ma}}, \bibinfo {author} {\bibfnamefont {X.}~\bibnamefont
  {Mi}}, \bibinfo {author} {\bibfnamefont {S.}~\bibnamefont {Misra}}, \bibinfo
  {author} {\bibfnamefont {C.}~\bibnamefont {Monroe}}, \bibinfo {author}
  {\bibfnamefont {K.}~\bibnamefont {Murch}}, \bibinfo {author} {\bibfnamefont
  {Z.}~\bibnamefont {Nazario}}, \bibinfo {author} {\bibfnamefont {K.-K.}\
  \bibnamefont {Ni}}, \bibinfo {author} {\bibfnamefont {A.~C.}\ \bibnamefont
  {Potter}}, \bibinfo {author} {\bibfnamefont {P.}~\bibnamefont {Roushan}},
  \bibinfo {author} {\bibfnamefont {M.}~\bibnamefont {Saffman}}, \bibinfo
  {author} {\bibfnamefont {M.}~\bibnamefont {Schleier-Smith}}, \bibinfo
  {author} {\bibfnamefont {I.}~\bibnamefont {Siddiqi}}, \bibinfo {author}
  {\bibfnamefont {R.}~\bibnamefont {Simmonds}}, \bibinfo {author}
  {\bibfnamefont {M.}~\bibnamefont {Singh}}, \bibinfo {author} {\bibfnamefont
  {I.}~\bibnamefont {Spielman}}, \bibinfo {author} {\bibfnamefont
  {K.}~\bibnamefont {Temme}}, \bibinfo {author} {\bibfnamefont {D.~S.}\
  \bibnamefont {Weiss}}, \bibinfo {author} {\bibfnamefont {J.}~\bibnamefont
  {Vu\ifmmode \check{c}\else \v{c}\fi{}kovi\ifmmode~\acute{c}\else
  \'{c}\fi{}}}, \bibinfo {author} {\bibfnamefont {V.}~\bibnamefont
  {Vuleti\ifmmode~\acute{c}\else \'{c}\fi{}}}, \bibinfo {author} {\bibfnamefont
  {J.}~\bibnamefont {Ye}}, \ and\ \bibinfo {author} {\bibfnamefont
  {M.}~\bibnamefont {Zwierlein}},\ }\href {\doibase
  10.1103/PRXQuantum.2.017003} {\bibfield  {journal} {\bibinfo  {journal} {PRX
  Quantum}\ }\textbf {\bibinfo {volume} {2}},\ \bibinfo {pages} {017003}
  (\bibinfo {year} {2021})}\BibitemShut {NoStop}%
\bibitem [{\citenamefont {Bloch}\ \emph {et~al.}(2012)\citenamefont {Bloch},
  \citenamefont {Dalibard},\ and\ \citenamefont {Nascimbene}}]{Bloch2012}%
  \BibitemOpen
  \bibfield  {author} {\bibinfo {author} {\bibfnamefont {I.}~\bibnamefont
  {Bloch}}, \bibinfo {author} {\bibfnamefont {J.}~\bibnamefont {Dalibard}}, \
  and\ \bibinfo {author} {\bibfnamefont {S.}~\bibnamefont {Nascimbene}},\
  }\href@noop {} {\bibfield  {journal} {\bibinfo  {journal} {Nature Physics}\
  }\textbf {\bibinfo {volume} {8}},\ \bibinfo {pages} {267} (\bibinfo {year}
  {2012})}\BibitemShut {NoStop}%
\bibitem [{\citenamefont {Bernien}\ \emph {et~al.}(2017)\citenamefont
  {Bernien}, \citenamefont {Schwartz}, \citenamefont {Keesling}, \citenamefont
  {Levine}, \citenamefont {Omran}, \citenamefont {Pichler}, \citenamefont
  {Choi}, \citenamefont {Zibrov}, \citenamefont {Endres}, \citenamefont
  {Greiner}, \citenamefont {Vuleti{\'{c}}},\ and\ \citenamefont
  {Lukin}}]{Bernien2017}%
  \BibitemOpen
  \bibfield  {author} {\bibinfo {author} {\bibfnamefont {H.}~\bibnamefont
  {Bernien}}, \bibinfo {author} {\bibfnamefont {S.}~\bibnamefont {Schwartz}},
  \bibinfo {author} {\bibfnamefont {A.}~\bibnamefont {Keesling}}, \bibinfo
  {author} {\bibfnamefont {H.}~\bibnamefont {Levine}}, \bibinfo {author}
  {\bibfnamefont {A.}~\bibnamefont {Omran}}, \bibinfo {author} {\bibfnamefont
  {H.}~\bibnamefont {Pichler}}, \bibinfo {author} {\bibfnamefont
  {S.}~\bibnamefont {Choi}}, \bibinfo {author} {\bibfnamefont {A.~S.}\
  \bibnamefont {Zibrov}}, \bibinfo {author} {\bibfnamefont {M.}~\bibnamefont
  {Endres}}, \bibinfo {author} {\bibfnamefont {M.}~\bibnamefont {Greiner}},
  \bibinfo {author} {\bibfnamefont {V.}~\bibnamefont {Vuleti{\'{c}}}}, \ and\
  \bibinfo {author} {\bibfnamefont {M.~D.}\ \bibnamefont {Lukin}},\ }\href
  {\doibase 10.1038/nature24622} {\bibfield  {journal} {\bibinfo  {journal}
  {Nature}\ }\textbf {\bibinfo {volume} {551}},\ \bibinfo {pages} {579}
  (\bibinfo {year} {2017})}\BibitemShut {NoStop}%
\bibitem [{\citenamefont {Bissbort}\ \emph {et~al.}(2013)\citenamefont
  {Bissbort}, \citenamefont {Cocks}, \citenamefont {Negretti}, \citenamefont
  {Idziaszek}, \citenamefont {Calarco}, \citenamefont {Schmidt-Kaler},
  \citenamefont {Hofstetter},\ and\ \citenamefont {Gerritsma}}]{Bissbort2013}%
  \BibitemOpen
  \bibfield  {author} {\bibinfo {author} {\bibfnamefont {U.}~\bibnamefont
  {Bissbort}}, \bibinfo {author} {\bibfnamefont {D.}~\bibnamefont {Cocks}},
  \bibinfo {author} {\bibfnamefont {A.}~\bibnamefont {Negretti}}, \bibinfo
  {author} {\bibfnamefont {Z.}~\bibnamefont {Idziaszek}}, \bibinfo {author}
  {\bibfnamefont {T.}~\bibnamefont {Calarco}}, \bibinfo {author} {\bibfnamefont
  {F.}~\bibnamefont {Schmidt-Kaler}}, \bibinfo {author} {\bibfnamefont
  {W.}~\bibnamefont {Hofstetter}}, \ and\ \bibinfo {author} {\bibfnamefont
  {R.}~\bibnamefont {Gerritsma}},\ }\href@noop {} {\bibfield  {journal}
  {\bibinfo  {journal} {Phys. Rev. Lett.}\ }\textbf {\bibinfo {volume} {111}},\
  \bibinfo {pages} {080501} (\bibinfo {year} {2013})}\BibitemShut {NoStop}%
\bibitem [{\citenamefont {Alexandrov}\ and\ \citenamefont
  {Ranninger}(1981)}]{Alexandrov1981}%
  \BibitemOpen
  \bibfield  {author} {\bibinfo {author} {\bibfnamefont {A.}~\bibnamefont
  {Alexandrov}}\ and\ \bibinfo {author} {\bibfnamefont {J.}~\bibnamefont
  {Ranninger}},\ }\href@noop {} {\bibfield  {journal} {\bibinfo  {journal}
  {Physical Review B}\ }\textbf {\bibinfo {volume} {24}},\ \bibinfo {pages}
  {1164} (\bibinfo {year} {1981})}\BibitemShut {NoStop}%
\bibitem [{\citenamefont {Lanzara}\ \emph {et~al.}(2001)\citenamefont
  {Lanzara}, \citenamefont {Bogdanov}, \citenamefont {Zhou}, \citenamefont
  {Kellar}, \citenamefont {Feng}, \citenamefont {Lu}, \citenamefont {Yoshida},
  \citenamefont {Eisaki}, \citenamefont {Fujimori}, \citenamefont {Kishio}
  \emph {et~al.}}]{lanzara2001evidence}%
  \BibitemOpen
  \bibfield  {author} {\bibinfo {author} {\bibfnamefont {A.}~\bibnamefont
  {Lanzara}}, \bibinfo {author} {\bibfnamefont {P.}~\bibnamefont {Bogdanov}},
  \bibinfo {author} {\bibfnamefont {X.}~\bibnamefont {Zhou}}, \bibinfo {author}
  {\bibfnamefont {S.}~\bibnamefont {Kellar}}, \bibinfo {author} {\bibfnamefont
  {D.}~\bibnamefont {Feng}}, \bibinfo {author} {\bibfnamefont {E.}~\bibnamefont
  {Lu}}, \bibinfo {author} {\bibfnamefont {T.}~\bibnamefont {Yoshida}},
  \bibinfo {author} {\bibfnamefont {H.}~\bibnamefont {Eisaki}}, \bibinfo
  {author} {\bibfnamefont {A.}~\bibnamefont {Fujimori}}, \bibinfo {author}
  {\bibfnamefont {K.}~\bibnamefont {Kishio}},  \emph {et~al.},\ }\href@noop {}
  {\bibfield  {journal} {\bibinfo  {journal} {Nature}\ }\textbf {\bibinfo
  {volume} {412}},\ \bibinfo {pages} {510} (\bibinfo {year}
  {2001})}\BibitemShut {NoStop}%
\bibitem [{\citenamefont {Pupillo}\ \emph {et~al.}(2008)\citenamefont
  {Pupillo}, \citenamefont {Griessner}, \citenamefont {Micheli}, \citenamefont
  {Ortner}, \citenamefont {Wang},\ and\ \citenamefont {Zoller}}]{Pupillo2008}%
  \BibitemOpen
  \bibfield  {author} {\bibinfo {author} {\bibfnamefont {G.}~\bibnamefont
  {Pupillo}}, \bibinfo {author} {\bibfnamefont {A.}~\bibnamefont {Griessner}},
  \bibinfo {author} {\bibfnamefont {A.}~\bibnamefont {Micheli}}, \bibinfo
  {author} {\bibfnamefont {M.}~\bibnamefont {Ortner}}, \bibinfo {author}
  {\bibfnamefont {D.-W.}\ \bibnamefont {Wang}}, \ and\ \bibinfo {author}
  {\bibfnamefont {P.}~\bibnamefont {Zoller}},\ }\href@noop {} {\bibfield
  {journal} {\bibinfo  {journal} {Phys. Rev. Lett.}\ }\textbf {\bibinfo
  {volume} {100}},\ \bibinfo {pages} {050402} (\bibinfo {year}
  {2008})}\BibitemShut {NoStop}%
\bibitem [{\citenamefont {Ortner}\ \emph {et~al.}(2009)\citenamefont {Ortner},
  \citenamefont {Micheli}, \citenamefont {Pupillo},\ and\ \citenamefont
  {Zoller}}]{Ortner2009}%
  \BibitemOpen
  \bibfield  {author} {\bibinfo {author} {\bibfnamefont {M.}~\bibnamefont
  {Ortner}}, \bibinfo {author} {\bibfnamefont {A.}~\bibnamefont {Micheli}},
  \bibinfo {author} {\bibfnamefont {G.}~\bibnamefont {Pupillo}}, \ and\
  \bibinfo {author} {\bibfnamefont {P.}~\bibnamefont {Zoller}},\ }\href@noop {}
  {\bibfield  {journal} {\bibinfo  {journal} {New Journal of Physics}\ }\textbf
  {\bibinfo {volume} {11}},\ \bibinfo {pages} {055045} (\bibinfo {year}
  {2009})}\BibitemShut {NoStop}%
\bibitem [{\citenamefont {Hague}\ and\ \citenamefont
  {MacCormick}(2012)}]{Hague2012}%
  \BibitemOpen
  \bibfield  {author} {\bibinfo {author} {\bibfnamefont {J.}~\bibnamefont
  {Hague}}\ and\ \bibinfo {author} {\bibfnamefont {C.}~\bibnamefont
  {MacCormick}},\ }\href@noop {} {\bibfield  {journal} {\bibinfo  {journal}
  {New Journal of Physics}\ }\textbf {\bibinfo {volume} {14}},\ \bibinfo
  {pages} {033019} (\bibinfo {year} {2012})}\BibitemShut {NoStop}%
\bibitem [{\citenamefont {Jachymski}\ and\ \citenamefont
  {Negretti}(2020)}]{Jachymski2020}%
  \BibitemOpen
  \bibfield  {author} {\bibinfo {author} {\bibfnamefont {K.}~\bibnamefont
  {Jachymski}}\ and\ \bibinfo {author} {\bibfnamefont {A.}~\bibnamefont
  {Negretti}},\ }\href {\doibase 10.1103/PhysRevResearch.2.033326} {\bibfield
  {journal} {\bibinfo  {journal} {Phys. Rev. Research}\ }\textbf {\bibinfo
  {volume} {2}},\ \bibinfo {pages} {033326} (\bibinfo {year}
  {2020})}\BibitemShut {NoStop}%
\bibitem [{\citenamefont {Wilson}\ \emph {et~al.}(2022)\citenamefont {Wilson},
  \citenamefont {Saskin}, \citenamefont {Meng}, \citenamefont {Ma},
  \citenamefont {Dilip}, \citenamefont {Burgers},\ and\ \citenamefont
  {Thompson}}]{Wilson2022}%
  \BibitemOpen
  \bibfield  {author} {\bibinfo {author} {\bibfnamefont {J.~T.}\ \bibnamefont
  {Wilson}}, \bibinfo {author} {\bibfnamefont {S.}~\bibnamefont {Saskin}},
  \bibinfo {author} {\bibfnamefont {Y.}~\bibnamefont {Meng}}, \bibinfo {author}
  {\bibfnamefont {S.}~\bibnamefont {Ma}}, \bibinfo {author} {\bibfnamefont
  {R.}~\bibnamefont {Dilip}}, \bibinfo {author} {\bibfnamefont {A.~P.}\
  \bibnamefont {Burgers}}, \ and\ \bibinfo {author} {\bibfnamefont {J.~D.}\
  \bibnamefont {Thompson}},\ }\href {\doibase 10.1103/PhysRevLett.128.033201}
  {\bibfield  {journal} {\bibinfo  {journal} {Phys. Rev. Lett.}\ }\textbf
  {\bibinfo {volume} {128}},\ \bibinfo {pages} {033201} (\bibinfo {year}
  {2022})}\BibitemShut {NoStop}%
\bibitem [{\citenamefont {Mei}\ \emph {et~al.}(2022)\citenamefont {Mei},
  \citenamefont {Li}, \citenamefont {Nguyen}, \citenamefont {Berman},\ and\
  \citenamefont {Kuzmich}}]{Kuzmich2022}%
  \BibitemOpen
  \bibfield  {author} {\bibinfo {author} {\bibfnamefont {Y.}~\bibnamefont
  {Mei}}, \bibinfo {author} {\bibfnamefont {Y.}~\bibnamefont {Li}}, \bibinfo
  {author} {\bibfnamefont {H.}~\bibnamefont {Nguyen}}, \bibinfo {author}
  {\bibfnamefont {P.~R.}\ \bibnamefont {Berman}}, \ and\ \bibinfo {author}
  {\bibfnamefont {A.}~\bibnamefont {Kuzmich}},\ }\href {\doibase
  10.1103/PhysRevLett.128.123601} {\bibfield  {journal} {\bibinfo  {journal}
  {Phys. Rev. Lett.}\ }\textbf {\bibinfo {volume} {128}},\ \bibinfo {pages}
  {123601} (\bibinfo {year} {2022})}\BibitemShut {NoStop}%
\bibitem [{\citenamefont {Lang}\ and\ \citenamefont {Firsov}(1963)}]{Lang1963}%
  \BibitemOpen
  \bibfield  {author} {\bibinfo {author} {\bibfnamefont {I.}~\bibnamefont
  {Lang}}\ and\ \bibinfo {author} {\bibfnamefont {Y.~A.}\ \bibnamefont
  {Firsov}},\ }\href@noop {} {\bibfield  {journal} {\bibinfo  {journal} {Sov.
  Phys. JETP}\ }\textbf {\bibinfo {volume} {16}},\ \bibinfo {pages} {1301}
  (\bibinfo {year} {1963})}\BibitemShut {NoStop}%
\bibitem [{\citenamefont {Wellein}\ \emph {et~al.}(1996)\citenamefont
  {Wellein}, \citenamefont {R\"oder},\ and\ \citenamefont
  {Fehske}}]{Fehske1996}%
  \BibitemOpen
  \bibfield  {author} {\bibinfo {author} {\bibfnamefont {G.}~\bibnamefont
  {Wellein}}, \bibinfo {author} {\bibfnamefont {H.}~\bibnamefont {R\"oder}}, \
  and\ \bibinfo {author} {\bibfnamefont {H.}~\bibnamefont {Fehske}},\ }\href
  {\doibase 10.1103/PhysRevB.53.9666} {\bibfield  {journal} {\bibinfo
  {journal} {Phys. Rev. B}\ }\textbf {\bibinfo {volume} {53}},\ \bibinfo
  {pages} {9666} (\bibinfo {year} {1996})}\BibitemShut {NoStop}%
\bibitem [{\citenamefont {Takada}\ and\ \citenamefont
  {Chatterjee}(2003)}]{Takada2003}%
  \BibitemOpen
  \bibfield  {author} {\bibinfo {author} {\bibfnamefont {Y.}~\bibnamefont
  {Takada}}\ and\ \bibinfo {author} {\bibfnamefont {A.}~\bibnamefont
  {Chatterjee}},\ }\href {\doibase 10.1103/PhysRevB.67.081102} {\bibfield
  {journal} {\bibinfo  {journal} {Phys. Rev. B}\ }\textbf {\bibinfo {volume}
  {67}},\ \bibinfo {pages} {081102} (\bibinfo {year} {2003})}\BibitemShut
  {NoStop}%
\bibitem [{\citenamefont {Clay}\ and\ \citenamefont
  {Hardikar}(2005)}]{Clay2005}%
  \BibitemOpen
  \bibfield  {author} {\bibinfo {author} {\bibfnamefont {R.~T.}\ \bibnamefont
  {Clay}}\ and\ \bibinfo {author} {\bibfnamefont {R.~P.}\ \bibnamefont
  {Hardikar}},\ }\href {\doibase 10.1103/PhysRevLett.95.096401} {\bibfield
  {journal} {\bibinfo  {journal} {Phys. Rev. Lett.}\ }\textbf {\bibinfo
  {volume} {95}},\ \bibinfo {pages} {096401} (\bibinfo {year}
  {2005})}\BibitemShut {NoStop}%
\bibitem [{\citenamefont {Hohenadler}\ and\ \citenamefont {von~der
  Linden}(2007)}]{Hohenadler2007}%
  \BibitemOpen
  \bibfield  {author} {\bibinfo {author} {\bibfnamefont {M.}~\bibnamefont
  {Hohenadler}}\ and\ \bibinfo {author} {\bibfnamefont {W.}~\bibnamefont
  {von~der Linden}},\ }in\ \href@noop {} {\emph {\bibinfo {booktitle} {Polarons
  in Advanced Materials}}}\ (\bibinfo  {publisher} {Springer},\ \bibinfo {year}
  {2007})\ pp.\ \bibinfo {pages} {463--502}\BibitemShut {NoStop}%
\bibitem [{\citenamefont {Tam}\ \emph {et~al.}(2014)\citenamefont {Tam},
  \citenamefont {Tsai},\ and\ \citenamefont {Campbell}}]{Campbell2014}%
  \BibitemOpen
  \bibfield  {author} {\bibinfo {author} {\bibfnamefont {K.-M.}\ \bibnamefont
  {Tam}}, \bibinfo {author} {\bibfnamefont {S.-W.}\ \bibnamefont {Tsai}}, \
  and\ \bibinfo {author} {\bibfnamefont {D.~K.}\ \bibnamefont {Campbell}},\
  }\href {\doibase 10.1103/PhysRevB.89.014513} {\bibfield  {journal} {\bibinfo
  {journal} {Phys. Rev. B}\ }\textbf {\bibinfo {volume} {89}},\ \bibinfo
  {pages} {014513} (\bibinfo {year} {2014})}\BibitemShut {NoStop}%
\bibitem [{\citenamefont {Yin}\ \emph {et~al.}(2015)\citenamefont {Yin},
  \citenamefont {Cocks},\ and\ \citenamefont {Hofstetter}}]{Yin2015}%
  \BibitemOpen
  \bibfield  {author} {\bibinfo {author} {\bibfnamefont {T.}~\bibnamefont
  {Yin}}, \bibinfo {author} {\bibfnamefont {D.}~\bibnamefont {Cocks}}, \ and\
  \bibinfo {author} {\bibfnamefont {W.}~\bibnamefont {Hofstetter}},\ }\href
  {\doibase 10.1103/PhysRevA.92.063635} {\bibfield  {journal} {\bibinfo
  {journal} {Phys. Rev. A}\ }\textbf {\bibinfo {volume} {92}},\ \bibinfo
  {pages} {063635} (\bibinfo {year} {2015})}\BibitemShut {NoStop}%
\bibitem [{\citenamefont {Wang}\ \emph {et~al.}(2020)\citenamefont {Wang},
  \citenamefont {Esterlis}, \citenamefont {Shi}, \citenamefont {Cirac},\ and\
  \citenamefont {Demler}}]{Wang2020}%
  \BibitemOpen
  \bibfield  {author} {\bibinfo {author} {\bibfnamefont {Y.}~\bibnamefont
  {Wang}}, \bibinfo {author} {\bibfnamefont {I.}~\bibnamefont {Esterlis}},
  \bibinfo {author} {\bibfnamefont {T.}~\bibnamefont {Shi}}, \bibinfo {author}
  {\bibfnamefont {J.~I.}\ \bibnamefont {Cirac}}, \ and\ \bibinfo {author}
  {\bibfnamefont {E.}~\bibnamefont {Demler}},\ }\href {\doibase
  10.1103/PhysRevResearch.2.043258} {\bibfield  {journal} {\bibinfo  {journal}
  {Phys. Rev. Research}\ }\textbf {\bibinfo {volume} {2}},\ \bibinfo {pages}
  {043258} (\bibinfo {year} {2020})}\BibitemShut {NoStop}%
\bibitem [{\citenamefont {de~Léséleuc}\ \emph {et~al.}(2019)\citenamefont
  {de~Léséleuc}, \citenamefont {Lienhard}, \citenamefont {Scholl},
  \citenamefont {Barredo}, \citenamefont {Weber}, \citenamefont {Lang},
  \citenamefont {Büchler}, \citenamefont {Lahaye},\ and\ \citenamefont
  {Browaeys}}]{Leseleuc2019}%
  \BibitemOpen
  \bibfield  {author} {\bibinfo {author} {\bibfnamefont {S.}~\bibnamefont
  {de~Léséleuc}}, \bibinfo {author} {\bibfnamefont {V.}~\bibnamefont
  {Lienhard}}, \bibinfo {author} {\bibfnamefont {P.}~\bibnamefont {Scholl}},
  \bibinfo {author} {\bibfnamefont {D.}~\bibnamefont {Barredo}}, \bibinfo
  {author} {\bibfnamefont {S.}~\bibnamefont {Weber}}, \bibinfo {author}
  {\bibfnamefont {N.}~\bibnamefont {Lang}}, \bibinfo {author} {\bibfnamefont
  {H.~P.}\ \bibnamefont {Büchler}}, \bibinfo {author} {\bibfnamefont
  {T.}~\bibnamefont {Lahaye}}, \ and\ \bibinfo {author} {\bibfnamefont
  {A.}~\bibnamefont {Browaeys}},\ }\href {\doibase 10.1126/science.aav9105}
  {\bibfield  {journal} {\bibinfo  {journal} {Science}\ }\textbf {\bibinfo
  {volume} {365}},\ \bibinfo {pages} {775} (\bibinfo {year}
  {2019})}\BibitemShut {NoStop}%
\bibitem [{\citenamefont {Sompet}\ \emph {et~al.}(2022)\citenamefont {Sompet},
  \citenamefont {Hirthe}, \citenamefont {Bourgund}, \citenamefont {Chalopin},
  \citenamefont {Bibo}, \citenamefont {Koepsell}, \citenamefont {Bojovi{\'c}},
  \citenamefont {Verresen}, \citenamefont {Pollmann}, \citenamefont {Salomon}
  \emph {et~al.}}]{Sompet2022}%
  \BibitemOpen
  \bibfield  {author} {\bibinfo {author} {\bibfnamefont {P.}~\bibnamefont
  {Sompet}}, \bibinfo {author} {\bibfnamefont {S.}~\bibnamefont {Hirthe}},
  \bibinfo {author} {\bibfnamefont {D.}~\bibnamefont {Bourgund}}, \bibinfo
  {author} {\bibfnamefont {T.}~\bibnamefont {Chalopin}}, \bibinfo {author}
  {\bibfnamefont {J.}~\bibnamefont {Bibo}}, \bibinfo {author} {\bibfnamefont
  {J.}~\bibnamefont {Koepsell}}, \bibinfo {author} {\bibfnamefont
  {P.}~\bibnamefont {Bojovi{\'c}}}, \bibinfo {author} {\bibfnamefont
  {R.}~\bibnamefont {Verresen}}, \bibinfo {author} {\bibfnamefont
  {F.}~\bibnamefont {Pollmann}}, \bibinfo {author} {\bibfnamefont
  {G.}~\bibnamefont {Salomon}},  \emph {et~al.},\ }\href@noop {} {\bibfield
  {journal} {\bibinfo  {journal} {Nature}\ ,\ \bibinfo {pages} {1}} (\bibinfo
  {year} {2022})}\BibitemShut {NoStop}%
\bibitem [{\citenamefont {Weber}\ \emph {et~al.}(2017)\citenamefont {Weber},
  \citenamefont {Tresp}, \citenamefont {Menke}, \citenamefont {Urvoy},
  \citenamefont {Firstenberg}, \citenamefont {B{\"u}chler},\ and\ \citenamefont
  {Hofferberth}}]{Weber2017}%
  \BibitemOpen
  \bibfield  {author} {\bibinfo {author} {\bibfnamefont {S.}~\bibnamefont
  {Weber}}, \bibinfo {author} {\bibfnamefont {C.}~\bibnamefont {Tresp}},
  \bibinfo {author} {\bibfnamefont {H.}~\bibnamefont {Menke}}, \bibinfo
  {author} {\bibfnamefont {A.}~\bibnamefont {Urvoy}}, \bibinfo {author}
  {\bibfnamefont {O.}~\bibnamefont {Firstenberg}}, \bibinfo {author}
  {\bibfnamefont {H.~P.}\ \bibnamefont {B{\"u}chler}}, \ and\ \bibinfo {author}
  {\bibfnamefont {S.}~\bibnamefont {Hofferberth}},\ }\href@noop {} {\bibfield
  {journal} {\bibinfo  {journal} {Journal of Physics B: Atomic, Molecular and
  Optical Physics}\ }\textbf {\bibinfo {volume} {50}},\ \bibinfo {pages}
  {133001} (\bibinfo {year} {2017})}\BibitemShut {NoStop}%
\bibitem [{\citenamefont {Mahan}(2013)}]{mahan2013book}%
  \BibitemOpen
  \bibfield  {author} {\bibinfo {author} {\bibfnamefont {G.~D.}\ \bibnamefont
  {Mahan}},\ }\href@noop {} {\emph {\bibinfo {title} {Many-particle physics}}}\
  (\bibinfo  {publisher} {Springer Science \& Business Media},\ \bibinfo {year}
  {2013})\BibitemShut {NoStop}%
\bibitem [{Sup()}]{SupMat}%
  \BibitemOpen
  \href@noop {} {}\bibinfo {note} {See Supplemental Material at [URL will be
  inserted by publisher] for more details about the phonon band structure of
  the two-leg ladder and the derivation of the atom-phonon coupling
  term.}\BibitemShut {Stop}%
\bibitem [{\citenamefont {Bissbort}\ \emph {et~al.}(2016)\citenamefont
  {Bissbort}, \citenamefont {Hofstetter},\ and\ \citenamefont
  {Poletti}}]{bissbort2016}%
  \BibitemOpen
  \bibfield  {author} {\bibinfo {author} {\bibfnamefont {U.}~\bibnamefont
  {Bissbort}}, \bibinfo {author} {\bibfnamefont {W.}~\bibnamefont
  {Hofstetter}}, \ and\ \bibinfo {author} {\bibfnamefont {D.}~\bibnamefont
  {Poletti}},\ }\href {\doibase 10.1103/PhysRevB.94.214305} {\bibfield
  {journal} {\bibinfo  {journal} {Phys. Rev. B}\ }\textbf {\bibinfo {volume}
  {94}},\ \bibinfo {pages} {214305} (\bibinfo {year} {2016})}\BibitemShut
  {NoStop}%
\bibitem [{\citenamefont {Gambetta}\ \emph {et~al.}(2020)\citenamefont
  {Gambetta}, \citenamefont {Li}, \citenamefont {Schmidt-Kaler},\ and\
  \citenamefont {Lesanovsky}}]{Gambetta2020}%
  \BibitemOpen
  \bibfield  {author} {\bibinfo {author} {\bibfnamefont {F.~M.}\ \bibnamefont
  {Gambetta}}, \bibinfo {author} {\bibfnamefont {W.}~\bibnamefont {Li}},
  \bibinfo {author} {\bibfnamefont {F.}~\bibnamefont {Schmidt-Kaler}}, \ and\
  \bibinfo {author} {\bibfnamefont {I.}~\bibnamefont {Lesanovsky}},\ }\href
  {\doibase 10.1103/PhysRevLett.124.043402} {\bibfield  {journal} {\bibinfo
  {journal} {Phys. Rev. Lett.}\ }\textbf {\bibinfo {volume} {124}},\ \bibinfo
  {pages} {043402} (\bibinfo {year} {2020})}\BibitemShut {NoStop}%
\bibitem [{\citenamefont {Marzari}\ and\ \citenamefont
  {Vanderbilt}(1997)}]{Marzari1997}%
  \BibitemOpen
  \bibfield  {author} {\bibinfo {author} {\bibfnamefont {N.}~\bibnamefont
  {Marzari}}\ and\ \bibinfo {author} {\bibfnamefont {D.}~\bibnamefont
  {Vanderbilt}},\ }\href {\doibase 10.1103/PhysRevB.56.12847} {\bibfield
  {journal} {\bibinfo  {journal} {Phys. Rev. B}\ }\textbf {\bibinfo {volume}
  {56}},\ \bibinfo {pages} {12847} (\bibinfo {year} {1997})}\BibitemShut
  {NoStop}%
\bibitem [{\citenamefont {Ganczarek}\ \emph {et~al.}(2014)\citenamefont
  {Ganczarek}, \citenamefont {Modugno}, \citenamefont {Pettini},\ and\
  \citenamefont {Zakrzewski}}]{Ganczarek2014}%
  \BibitemOpen
  \bibfield  {author} {\bibinfo {author} {\bibfnamefont {W.}~\bibnamefont
  {Ganczarek}}, \bibinfo {author} {\bibfnamefont {M.}~\bibnamefont {Modugno}},
  \bibinfo {author} {\bibfnamefont {G.}~\bibnamefont {Pettini}}, \ and\
  \bibinfo {author} {\bibfnamefont {J.}~\bibnamefont {Zakrzewski}},\ }\href
  {\doibase 10.1103/PhysRevA.90.033621} {\bibfield  {journal} {\bibinfo
  {journal} {Phys. Rev. A}\ }\textbf {\bibinfo {volume} {90}},\ \bibinfo
  {pages} {033621} (\bibinfo {year} {2014})}\BibitemShut {NoStop}%
\bibitem [{\citenamefont {Negretti}\ \emph {et~al.}(2014)\citenamefont
  {Negretti}, \citenamefont {Gerritsma}, \citenamefont {Idziaszek},
  \citenamefont {Schmidt-Kaler},\ and\ \citenamefont {Calarco}}]{Negretti2014}%
  \BibitemOpen
  \bibfield  {author} {\bibinfo {author} {\bibfnamefont {A.}~\bibnamefont
  {Negretti}}, \bibinfo {author} {\bibfnamefont {R.}~\bibnamefont {Gerritsma}},
  \bibinfo {author} {\bibfnamefont {Z.}~\bibnamefont {Idziaszek}}, \bibinfo
  {author} {\bibfnamefont {F.}~\bibnamefont {Schmidt-Kaler}}, \ and\ \bibinfo
  {author} {\bibfnamefont {T.}~\bibnamefont {Calarco}},\ }\href {\doibase
  10.1103/PhysRevB.90.155426} {\bibfield  {journal} {\bibinfo  {journal} {Phys.
  Rev. B}\ }\textbf {\bibinfo {volume} {90}},\ \bibinfo {pages} {155426}
  (\bibinfo {year} {2014})}\BibitemShut {NoStop}%
\bibitem [{\citenamefont {Kokail}\ \emph {et~al.}(2019)\citenamefont {Kokail},
  \citenamefont {Maier}, \citenamefont {van Bijnen}, \citenamefont {Brydges},
  \citenamefont {Joshi}, \citenamefont {Jurcevic}, \citenamefont {Muschik},
  \citenamefont {Silvi}, \citenamefont {Blatt}, \citenamefont {Roos} \emph
  {et~al.}}]{Kokail2019}%
  \BibitemOpen
  \bibfield  {author} {\bibinfo {author} {\bibfnamefont {C.}~\bibnamefont
  {Kokail}}, \bibinfo {author} {\bibfnamefont {C.}~\bibnamefont {Maier}},
  \bibinfo {author} {\bibfnamefont {R.}~\bibnamefont {van Bijnen}}, \bibinfo
  {author} {\bibfnamefont {T.}~\bibnamefont {Brydges}}, \bibinfo {author}
  {\bibfnamefont {M.~K.}\ \bibnamefont {Joshi}}, \bibinfo {author}
  {\bibfnamefont {P.}~\bibnamefont {Jurcevic}}, \bibinfo {author}
  {\bibfnamefont {C.~A.}\ \bibnamefont {Muschik}}, \bibinfo {author}
  {\bibfnamefont {P.}~\bibnamefont {Silvi}}, \bibinfo {author} {\bibfnamefont
  {R.}~\bibnamefont {Blatt}}, \bibinfo {author} {\bibfnamefont {C.~F.}\
  \bibnamefont {Roos}},  \emph {et~al.},\ }\href@noop {} {\bibfield  {journal}
  {\bibinfo  {journal} {Nature}\ }\textbf {\bibinfo {volume} {569}},\ \bibinfo
  {pages} {355} (\bibinfo {year} {2019})}\BibitemShut {NoStop}%
\bibitem [{\citenamefont {Meth}\ \emph {et~al.}(2022)\citenamefont {Meth},
  \citenamefont {Kuzmin}, \citenamefont {van Bijnen}, \citenamefont {Postler},
  \citenamefont {Stricker}, \citenamefont {Blatt}, \citenamefont {Ringbauer},
  \citenamefont {Monz}, \citenamefont {Silvi},\ and\ \citenamefont
  {Schindler}}]{Meth2022}%
  \BibitemOpen
  \bibfield  {author} {\bibinfo {author} {\bibfnamefont {M.}~\bibnamefont
  {Meth}}, \bibinfo {author} {\bibfnamefont {V.}~\bibnamefont {Kuzmin}},
  \bibinfo {author} {\bibfnamefont {R.}~\bibnamefont {van Bijnen}}, \bibinfo
  {author} {\bibfnamefont {L.}~\bibnamefont {Postler}}, \bibinfo {author}
  {\bibfnamefont {R.}~\bibnamefont {Stricker}}, \bibinfo {author}
  {\bibfnamefont {R.}~\bibnamefont {Blatt}}, \bibinfo {author} {\bibfnamefont
  {M.}~\bibnamefont {Ringbauer}}, \bibinfo {author} {\bibfnamefont
  {T.}~\bibnamefont {Monz}}, \bibinfo {author} {\bibfnamefont {P.}~\bibnamefont
  {Silvi}}, \ and\ \bibinfo {author} {\bibfnamefont {P.}~\bibnamefont
  {Schindler}},\ }\href@noop {} {\bibfield  {journal} {\bibinfo  {journal}
  {arXiv preprint arXiv:2203.13271}\ } (\bibinfo {year} {2022})}\BibitemShut
  {NoStop}%
\bibitem [{\citenamefont {Shi}\ and\ \citenamefont {Cirac}(2013)}]{Shi2013}%
  \BibitemOpen
  \bibfield  {author} {\bibinfo {author} {\bibfnamefont {T.}~\bibnamefont
  {Shi}}\ and\ \bibinfo {author} {\bibfnamefont {J.~I.}\ \bibnamefont
  {Cirac}},\ }\href {\doibase 10.1103/PhysRevA.87.013606} {\bibfield  {journal}
  {\bibinfo  {journal} {Phys. Rev. A}\ }\textbf {\bibinfo {volume} {87}},\
  \bibinfo {pages} {013606} (\bibinfo {year} {2013})}\BibitemShut {NoStop}%
\bibitem [{\citenamefont {Reshodko}\ \emph {et~al.}(2019)\citenamefont
  {Reshodko}, \citenamefont {Benseny}, \citenamefont {Romh{\'a}nyi},\ and\
  \citenamefont {Busch}}]{Reshodko2019}%
  \BibitemOpen
  \bibfield  {author} {\bibinfo {author} {\bibfnamefont {I.}~\bibnamefont
  {Reshodko}}, \bibinfo {author} {\bibfnamefont {A.}~\bibnamefont {Benseny}},
  \bibinfo {author} {\bibfnamefont {J.}~\bibnamefont {Romh{\'a}nyi}}, \ and\
  \bibinfo {author} {\bibfnamefont {T.}~\bibnamefont {Busch}},\ }\href@noop {}
  {\bibfield  {journal} {\bibinfo  {journal} {New Journal of Physics}\ }\textbf
  {\bibinfo {volume} {21}},\ \bibinfo {pages} {013010} (\bibinfo {year}
  {2019})}\BibitemShut {NoStop}%
\bibitem [{\citenamefont {Salamon}\ \emph {et~al.}(2020)\citenamefont
  {Salamon}, \citenamefont {Celi}, \citenamefont {Chhajlany}, \citenamefont
  {Fr\'erot}, \citenamefont {Lewenstein}, \citenamefont {Tarruell},\ and\
  \citenamefont {Rakshit}}]{Salamon2020}%
  \BibitemOpen
  \bibfield  {author} {\bibinfo {author} {\bibfnamefont {T.}~\bibnamefont
  {Salamon}}, \bibinfo {author} {\bibfnamefont {A.}~\bibnamefont {Celi}},
  \bibinfo {author} {\bibfnamefont {R.~W.}\ \bibnamefont {Chhajlany}}, \bibinfo
  {author} {\bibfnamefont {I.}~\bibnamefont {Fr\'erot}}, \bibinfo {author}
  {\bibfnamefont {M.}~\bibnamefont {Lewenstein}}, \bibinfo {author}
  {\bibfnamefont {L.}~\bibnamefont {Tarruell}}, \ and\ \bibinfo {author}
  {\bibfnamefont {D.}~\bibnamefont {Rakshit}},\ }\href {\doibase
  10.1103/PhysRevLett.125.030504} {\bibfield  {journal} {\bibinfo  {journal}
  {Phys. Rev. Lett.}\ }\textbf {\bibinfo {volume} {125}},\ \bibinfo {pages}
  {030504} (\bibinfo {year} {2020})}\BibitemShut {NoStop}%
\bibitem [{\citenamefont {Kn\"orzer}\ \emph {et~al.}(2022)\citenamefont
  {Kn\"orzer}, \citenamefont {Shi}, \citenamefont {Demler},\ and\ \citenamefont
  {Cirac}}]{Knorzer2022}%
  \BibitemOpen
  \bibfield  {author} {\bibinfo {author} {\bibfnamefont {J.}~\bibnamefont
  {Kn\"orzer}}, \bibinfo {author} {\bibfnamefont {T.}~\bibnamefont {Shi}},
  \bibinfo {author} {\bibfnamefont {E.}~\bibnamefont {Demler}}, \ and\ \bibinfo
  {author} {\bibfnamefont {J.~I.}\ \bibnamefont {Cirac}},\ }\href {\doibase
  10.1103/PhysRevLett.128.120404} {\bibfield  {journal} {\bibinfo  {journal}
  {Phys. Rev. Lett.}\ }\textbf {\bibinfo {volume} {128}},\ \bibinfo {pages}
  {120404} (\bibinfo {year} {2022})}\BibitemShut {NoStop}%
\bibitem [{\citenamefont {Di~Liberto}\ \emph {et~al.}(2022)\citenamefont
  {Di~Liberto}, \citenamefont {Kruckenhauser}, \citenamefont {Zoller},\ and\
  \citenamefont {Baranov}}]{DiLiberto2022}%
  \BibitemOpen
  \bibfield  {author} {\bibinfo {author} {\bibfnamefont {M.}~\bibnamefont
  {Di~Liberto}}, \bibinfo {author} {\bibfnamefont {A.}~\bibnamefont
  {Kruckenhauser}}, \bibinfo {author} {\bibfnamefont {P.}~\bibnamefont
  {Zoller}}, \ and\ \bibinfo {author} {\bibfnamefont {M.~A.}\ \bibnamefont
  {Baranov}},\ }\href {\doibase 10.22331/q-2022-06-07-731} {\bibfield
  {journal} {\bibinfo  {journal} {{Quantum}}\ }\textbf {\bibinfo {volume}
  {6}},\ \bibinfo {pages} {731} (\bibinfo {year} {2022})}\BibitemShut {NoStop}%
\bibitem [{\citenamefont {Cavalleri}(2018)}]{Cavalleri2018}%
  \BibitemOpen
  \bibfield  {author} {\bibinfo {author} {\bibfnamefont {A.}~\bibnamefont
  {Cavalleri}},\ }\href@noop {} {\bibfield  {journal} {\bibinfo  {journal}
  {Contemporary Physics}\ }\textbf {\bibinfo {volume} {59}},\ \bibinfo {pages}
  {31} (\bibinfo {year} {2018})}\BibitemShut {NoStop}%
\bibitem [{\citenamefont {Babadi}\ \emph {et~al.}(2017)\citenamefont {Babadi},
  \citenamefont {Knap}, \citenamefont {Martin}, \citenamefont {Refael},\ and\
  \citenamefont {Demler}}]{Babadi2017}%
  \BibitemOpen
  \bibfield  {author} {\bibinfo {author} {\bibfnamefont {M.}~\bibnamefont
  {Babadi}}, \bibinfo {author} {\bibfnamefont {M.}~\bibnamefont {Knap}},
  \bibinfo {author} {\bibfnamefont {I.}~\bibnamefont {Martin}}, \bibinfo
  {author} {\bibfnamefont {G.}~\bibnamefont {Refael}}, \ and\ \bibinfo {author}
  {\bibfnamefont {E.}~\bibnamefont {Demler}},\ }\href {\doibase
  10.1103/PhysRevB.96.014512} {\bibfield  {journal} {\bibinfo  {journal} {Phys.
  Rev. B}\ }\textbf {\bibinfo {volume} {96}},\ \bibinfo {pages} {014512}
  (\bibinfo {year} {2017})}\BibitemShut {NoStop}%
\end{thebibliography}%


\begin{thebibliography}{1}

\bibitem{mahan2013book}
Gerald~D Mahan.
\newblock {\em Many-particle physics}.
\newblock Springer Science \& Business Media, 2013.

\bibitem{Bissbort2016}
U.~Bissbort, W.~Hofstetter, and D.~Poletti.
\newblock Operator-based derivation of phonon modes and characterization of
  correlations for trapped ions at zero and finite temperature.
\newblock {\em Phys. Rev. B}, 94:214305, Dec 2016.

\bibitem{fetter2012quantum}
Alexander~L Fetter and John~Dirk Walecka.
\newblock {\em Quantum theory of many-particle systems}.
\newblock Courier Corporation, 2012.

\bibitem{Ortner2009}
M~Ortner, A~Micheli, G~Pupillo, and P~Zoller.
\newblock Quantum simulations of extended hubbard models with dipolar crystals.
\newblock {\em New Journal of Physics}, 11(5):055045, 2009.

\bibitem{Marzari1997}
Nicola Marzari and David Vanderbilt.
\newblock Maximally localized generalized wannier functions for composite
  energy bands.
\newblock {\em Phys. Rev. B}, 56:12847--12865, Nov 1997.

\bibitem{Ganczarek2014}
Wojciech Ganczarek, Michele Modugno, Giulio Pettini, and Jakub Zakrzewski.
\newblock Wannier functions for one-dimensional $s\ensuremath{-}p$ optical
  superlattices.
\newblock {\em Phys. Rev. A}, 90:033621, Sep 2014.

\bibitem{Negretti2014}
A.~Negretti, R.~Gerritsma, Z.~Idziaszek, F.~Schmidt-Kaler, and T.~Calarco.
\newblock Generalized kronig-penney model for ultracold atomic quantum systems.
\newblock {\em Phys. Rev. B}, 90:155426, Oct 2014.

\end{thebibliography}
\end{document}


\title{Supplemental Material to ``Quantum simulation of extended electron-phonon coupling models in a hybrid Rydberg atom setup''}
\author{Jo{\~{a}}o P. Mendon{\c{c}}a}
\email{jpedromend@gmail.com}
\affiliation{%
		Faculty of Physics, University of Warsaw, Pasteura 5, 02-093 Warsaw, Poland 
	}
\author{Krzysztof Jachymski}
\affiliation{%
		Faculty of Physics, University of Warsaw, Pasteura 5, 02-093 Warsaw, Poland 
	}
\date{\today}
\maketitle

\section{Phonon band structure}
The phonon properties were obtained through the first and second derivatives of the potential energy. Below we give some analytical details and important numerical results not shown in the main part. First, the potential energy is written as
\begin{align}
    V = \frac{1}{2}\sum_{n,\alpha}M_{\alpha}[\vb*{\nu}_n(\vb{R}_{n\alpha}-\Bar{\vb{R}}_{n\alpha})]^2 +\sum_{\substack{nm\alpha\beta\\(n,\alpha)\neq(m,\beta)}} \frac{V_{\textrm{dd}}}{|\vb{R}_{n\alpha,m\beta}|^3}\left[1 - 3(\vu{m}\cdot \vu{R}_{n\alpha,m\beta})^2\right],
\end{align}
where $\vu{R}_{n\alpha,m\beta}=\vu{R}_{n\alpha}-\vu{R}_{m\beta}$. We can write the potential in the harmonic approximation,
\begin{equation}
    V\approx V_0 + \frac{1}{2}\sum_{nm}\sum_{\alpha\beta}\sum_{ij} u_{n\alpha i}D_{n\alpha i}^{m\beta j}u_{m\beta j},
\end{equation}
where the harmonic matrix $D$ is given by the second derivatives at equilibrium,
$D_{n\alpha i}^{m\beta j} = (\pdv*{V}{R_{m\beta j}}{R_{n\alpha i}})\eval_{\textrm{eq}}$.
The equilibrium positions are given by $(\pdv*{V}{R_{n\alpha i}})\eval_{\textrm{eq}}=0$.
In a translationally invariant system, we can write $D_{\alpha i}^{\beta j}(|\vb{R}_n-\vb{R}_m|)=D_{\alpha i}^{\beta j}(\vb{R}_p)$. 
That way, the dynamical matrix elements are given by~\cite{mahan2013book},
\begin{equation}
    \Tilde{D}_{\alpha i}^{\beta j}(\vb{q}) = \sum_n D_{\alpha i}^{\beta j}(\vb{R}_n)e^{i\vb{q}\cdot\vb{R}_n},
\end{equation}
where $\vb{R}_{n\alpha}(t)=\vb{R}_n + \vb*{\rho}_{\alpha} + \vb{u}_{n\alpha}(t)$ and $\vb{R}_n=na\vu{z}$. Because of the quasi one-dimensional geometrical nature of our system the cell position vectors $\vb{R}_n$ point only on $z$--direction.

The properties of the band structure are very rich even in our simple quasi-1D model. There are band crossings, shape and width changes, and more features indicated below. Here in all cases we fixed $N=14,\Delta=1,a=2d,\theta=\theta_m,\phi=0$.
\subsection{Band crossing}
\begin{figure}[H]
    \centering
    \includegraphics[width=0.8\linewidth]{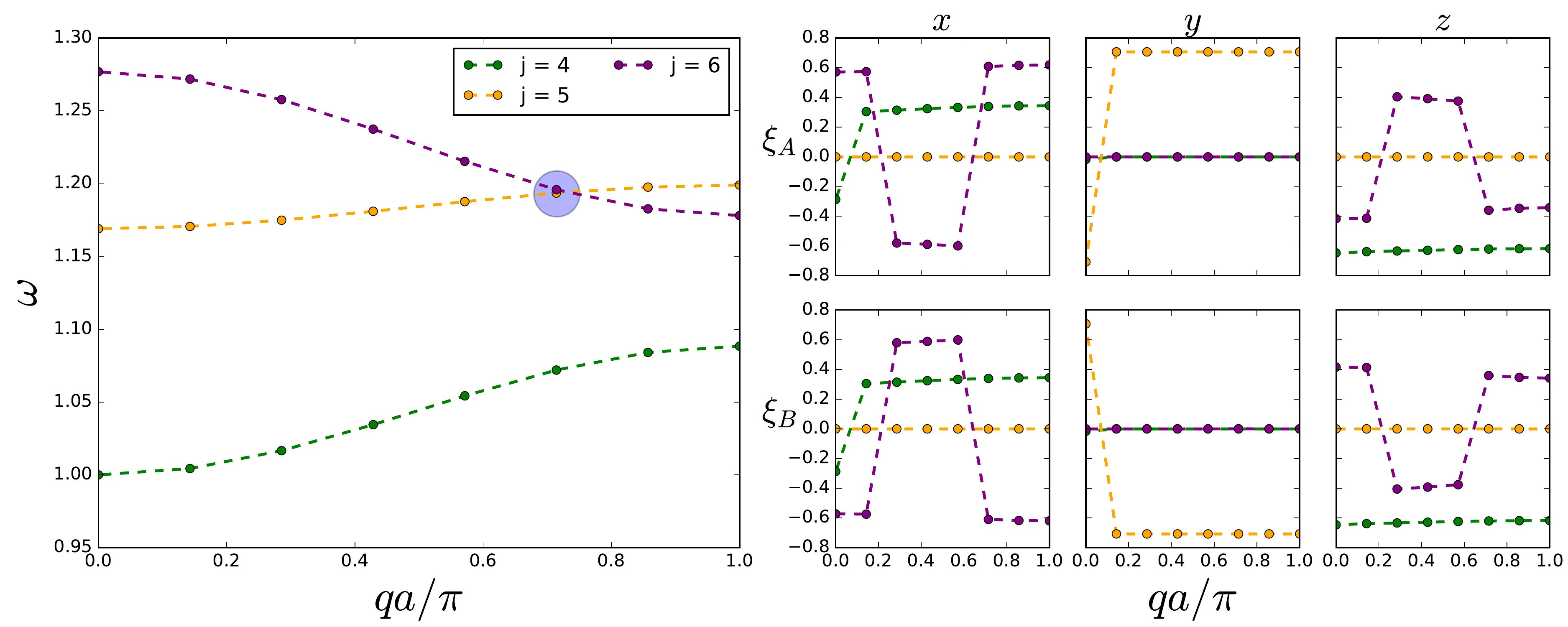}
    \caption{(left) Eigenvalues and (right) corresponding eigenvectors of the dynamical matrix for $d=1.5$. The bands $j=5$ and $j=6$ are crossing.}
    \label{fig:1}
\end{figure}
\begin{figure}[H]
    \centering
    \includegraphics[width=0.8\linewidth]{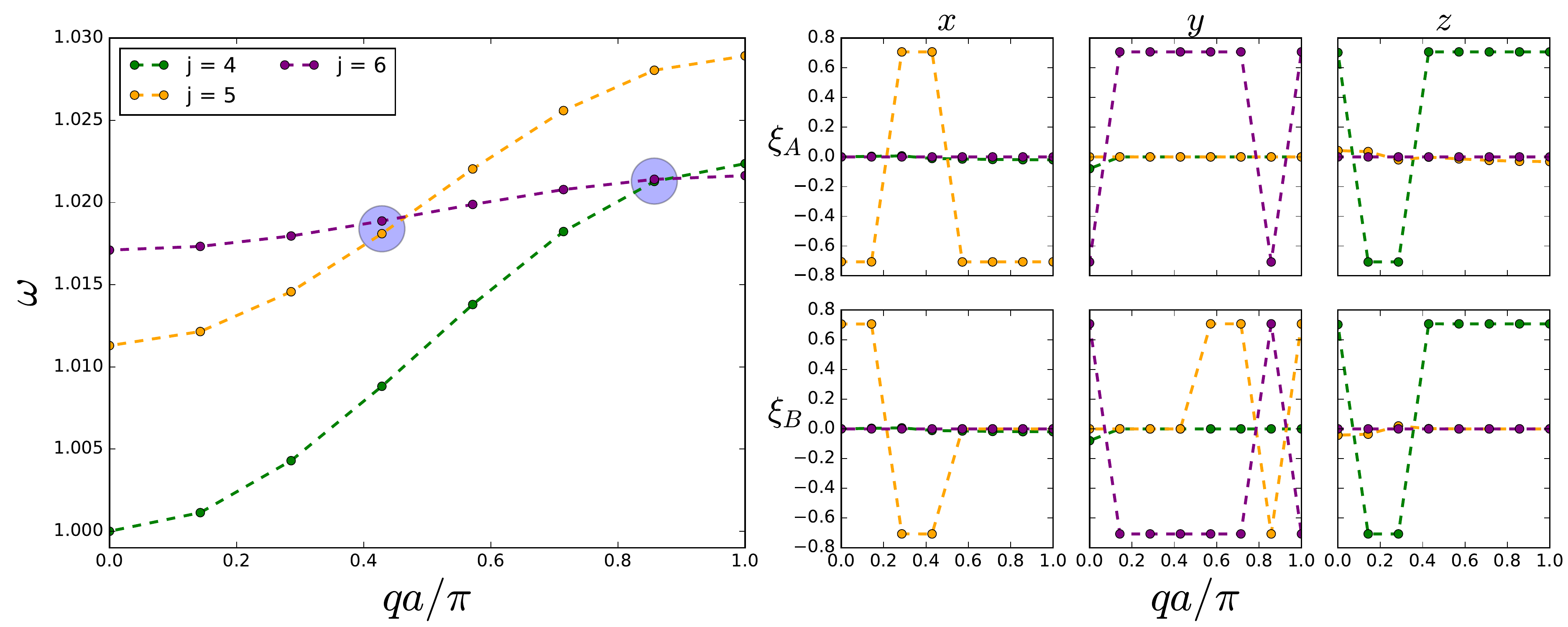}
    \caption{(left) Eigenvalues and (right) corresponding eigenvectors of the dynamical matrix for $d=2.5$. The highest band $j=6$ is crossing both $j=5$ and $j=4$ bands.}
    \label{fig:2}
\end{figure}

As observed in the letter, for both $d=1.5$ and $d=2.5$ some bands are crossing. Let us take a deeper look on that. In the Figures \ref{fig:1} and \ref{fig:2}, we show the three highest modes and marked with a blue circle the $q$ points where the band are crossing.
\subsection{Concavity change}
\begin{figure}[H]
    \centering
    \includegraphics[width=0.45\linewidth]{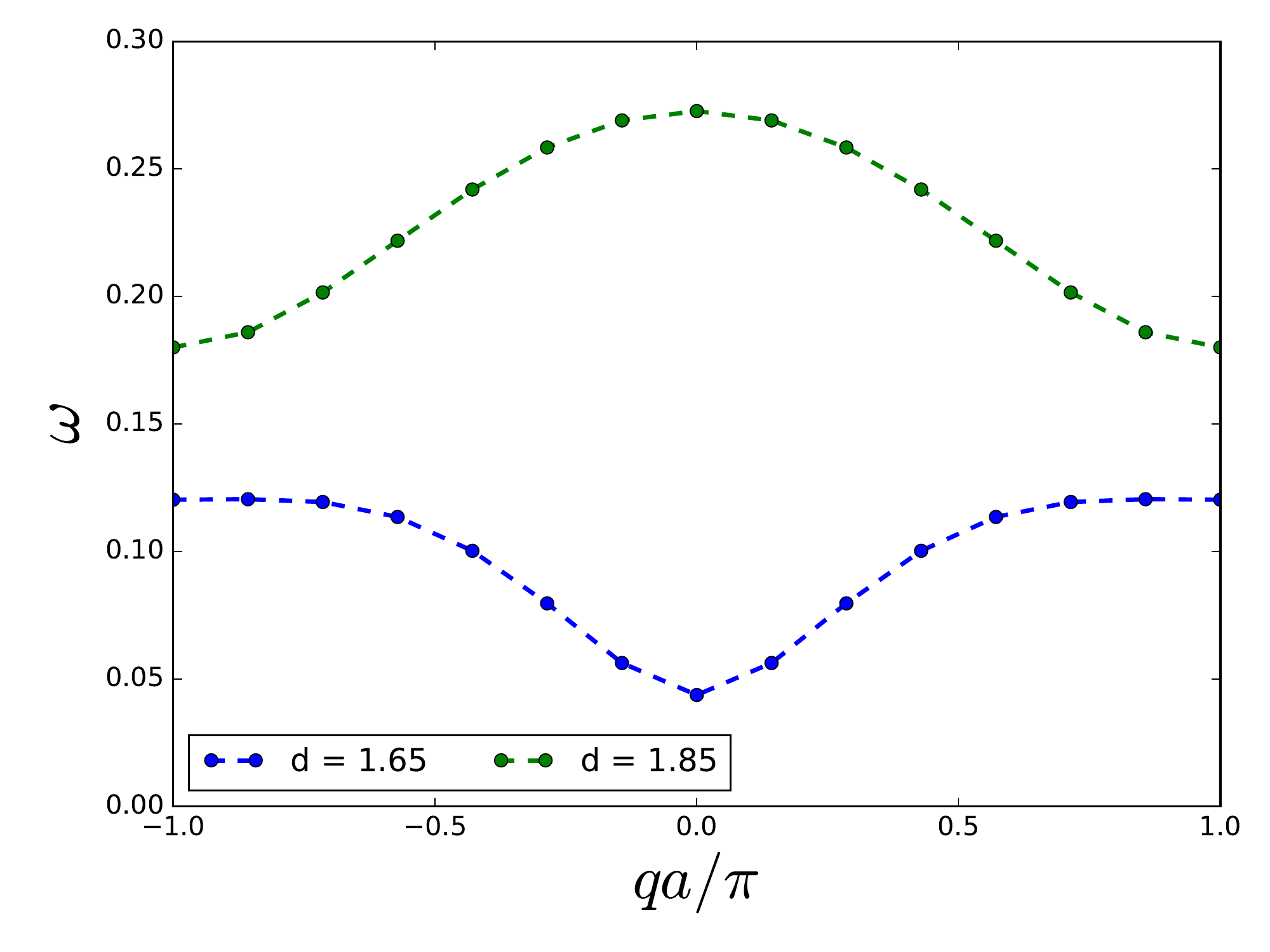}
    \includegraphics[width=0.45\linewidth]{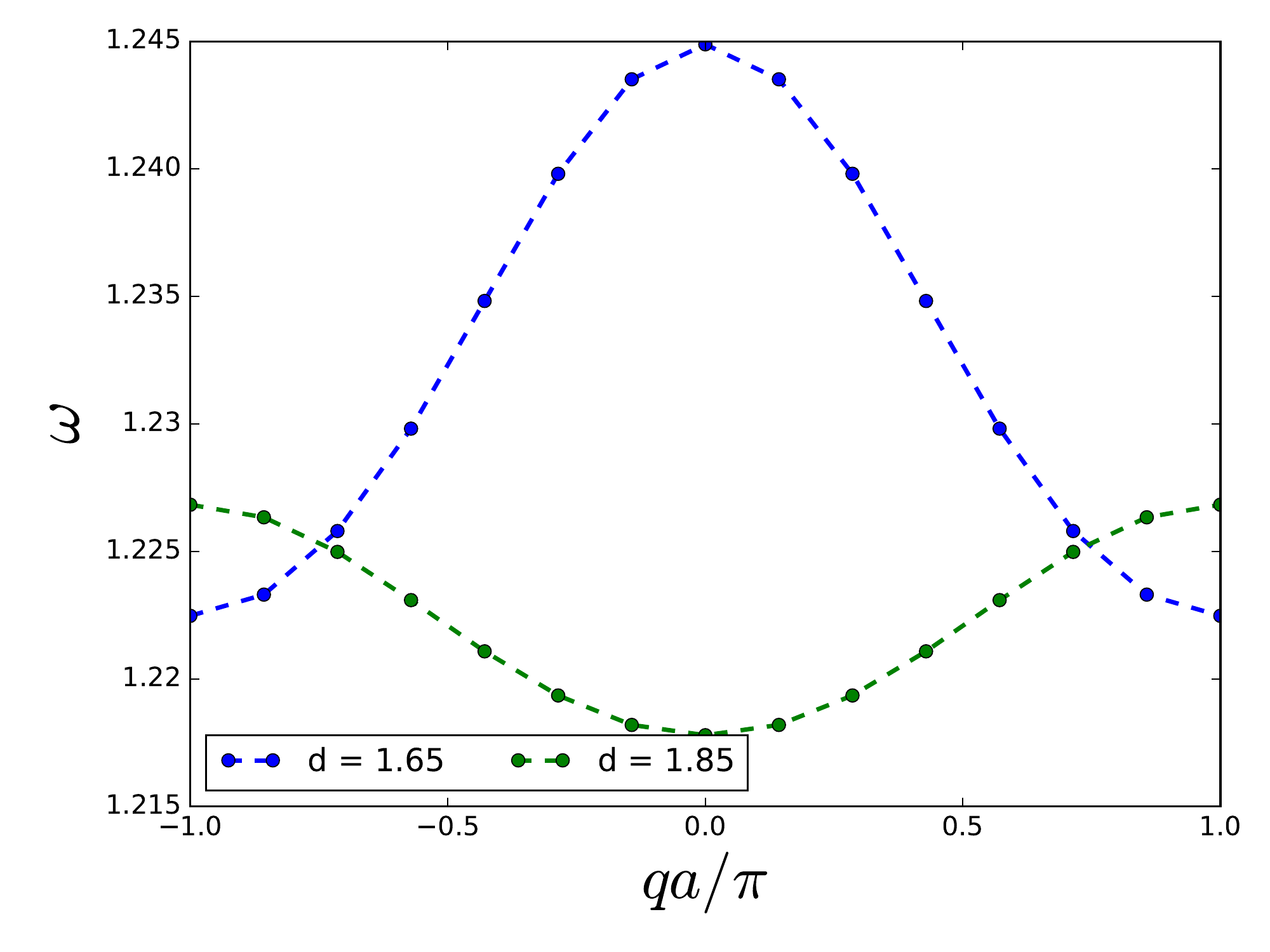}
    \caption{Phonon spectrum of the (left) 1st band and (right) 6th band for two different $d$ values.}
    \label{fig:3}
\end{figure}

Both highest and lowest bands are changing their concavity with varying the distance $d$. We show in Fig. \ref{fig:3} two examples where there is a change in concavity in both lowest and highest bands.
It is hard to find numerically the critical value at which the concavity changes, as one encounters instabilities.  As demonstrated in the figure, it is located between $d=1.65$ and $=1.85$.
\subsection{Non-monotonic 1st band interaction strength}
\begin{figure}[H]
    \centering
    \includegraphics[width=0.6\linewidth]{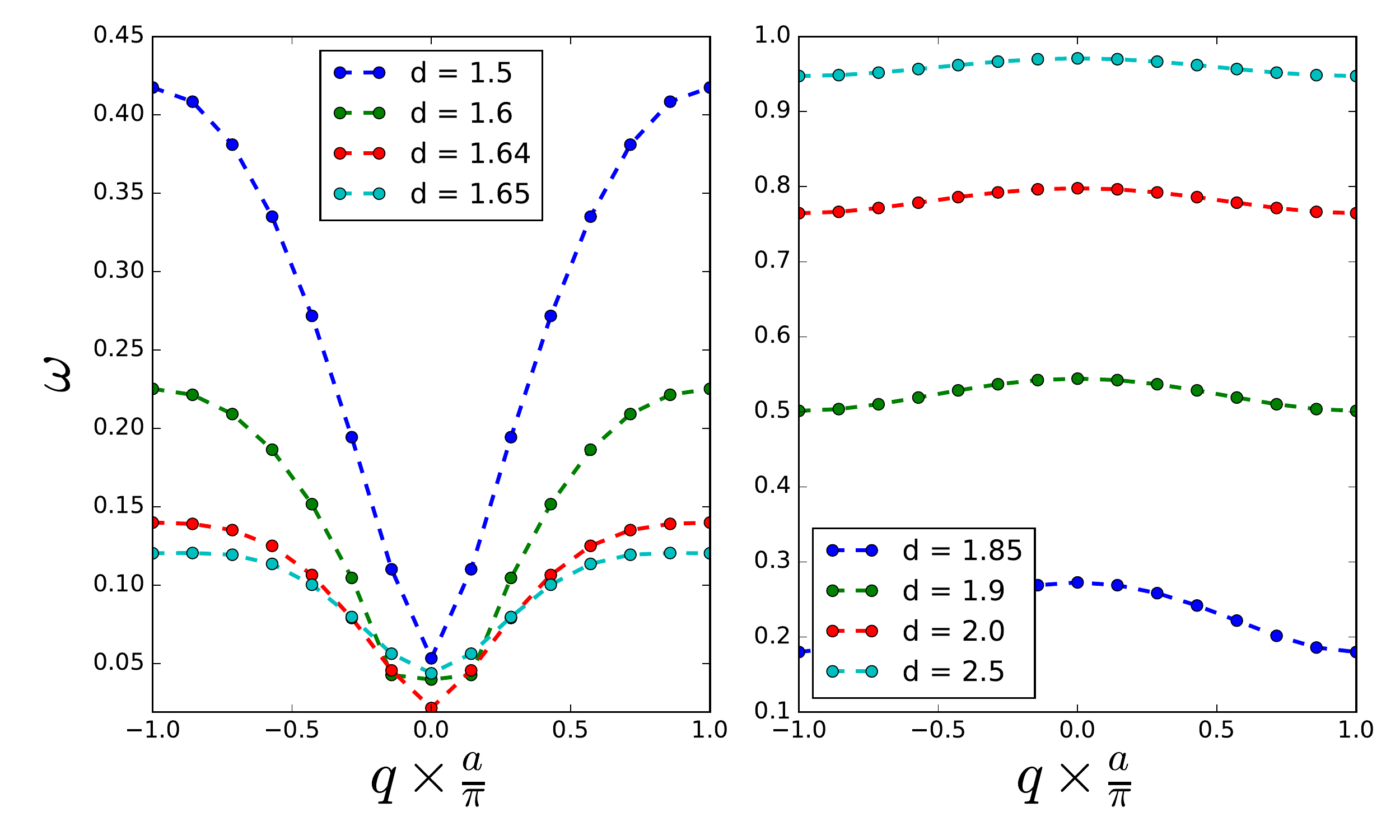}
    \caption{Phonon spectrum of the 1st band for many values of $d$.}
    \label{fig:4}
\end{figure}

The atom-phonon coupling strength to the first phonon band shows a non-monotonic behavior due to the non-monotonic behavior of $\omega_{j=1}$. As we can see in Fig.~\ref{fig:4}, from $d=1.5$ to $1.65$ the band is flattening, which increase the interaction strength due to the phonon energy being in the denominator. After the concavity changing transition, the band start to widen again. This strongly suggests that the first band is flat at the transition point.

\subsection{Local phonon interactions}
We can also move to the local phonon study, following Bissbort \textit{et al} \cite{Bissbort2016}, to obtain a good picture of the Rydberg--Rydberg interactions in the harmonic approximation. We can rewrite the phonon Hamiltonian as
\begin{align}\label{eq:Hbare}
    H &= \sum_{n,i} \frac{P_{n,i}^2}{2M_n} + \frac{1}{2}\sum_{n,m,i,j} u_{n,i} D_{nm}^{ij} u_{m,j} \nonumber\\
    &= \frac{1}{2}\sum_{n,i}\left( \frac{P_{n,i}^2}{M_n} + D_{nn}^{ii} u_{n,i}^2 \right) + \frac{1}{2}\sum_{\substack{n,m,i,j \\ (n,i) \neq (m,j)}} u_{n,i} D_{nm}^{ij} u_{m,j}.
\end{align}
We can second quantize it defining,
\begin{align}
    u_{n,i} &= \sqrt{\frac{1}{2M_n\Omega_{n,i}}}(b_{n,i}+b_{n,i}^{\dagger}), \\
    P_{n,i} &= i\sqrt{\frac{2M_n\Omega_{n,i}}{2}}(b_{n,i}^{\dagger}-b_{n,i}),
\end{align}
where $D_{nn}^{ii}=M_n\Omega_{n,i}^2$. The first term in Eq. \ref{eq:Hbare} is a collection of local harmonic oscillators $\frac{1}{2}\sum_{n,i}\Omega_{n,i}( b_{n,i}^{\dagger}b_{n,i} + b_{n,i}b_{n,i}^{\dagger})$. The second term represents the interaction part of the Hamiltonian,
\begin{align}
    \frac{1}{2}\sum_{\substack{n,m,i,j \\ (n,i) \neq (m,j)}} \frac{D_{nm}^{ij}}{2M_n\sqrt{\Omega_{n,i}\Omega_{m,j}}}( b_{n,i}b_{m,j} + b_{n,i}b_{m,j}^{\dagger} + b_{n,i}^{\dagger}b_{m,j} + b_{n,i}^{\dagger}b_{m,j}^{\dagger}).
\end{align}
The full phonon Hamiltonian can be written in a more convenient way,
\begin{align}
    H = \frac{1}{2}\sum_{nmij}\bigg[ h_{nm}^{ij}( b_{n,i}^{\dagger}b_{m,j} + b_{n,i}b_{m,j}^{\dagger})  + g_{nm}^{ij}( b_{n,i}b_{m,j} + b_{n,i}^{\dagger}b_{m,j}^{\dagger}) \bigg],
\end{align}
where 
\begin{align}
    g_{nm}^{ij} &= \frac{(1-\delta_{nm}\delta_{ij})D_{nm}^{ij}}{2M_n\sqrt{\Omega_{n,i}\Omega_{m,j}}}, \\
    h_{nm}^{ij} &= \delta_{nm}\delta_{ij}\Omega_{n,i} + g_{nm}^{ij}.
\end{align}

\begin{figure}[H]
    \centering
    \includegraphics[width=0.5\linewidth]{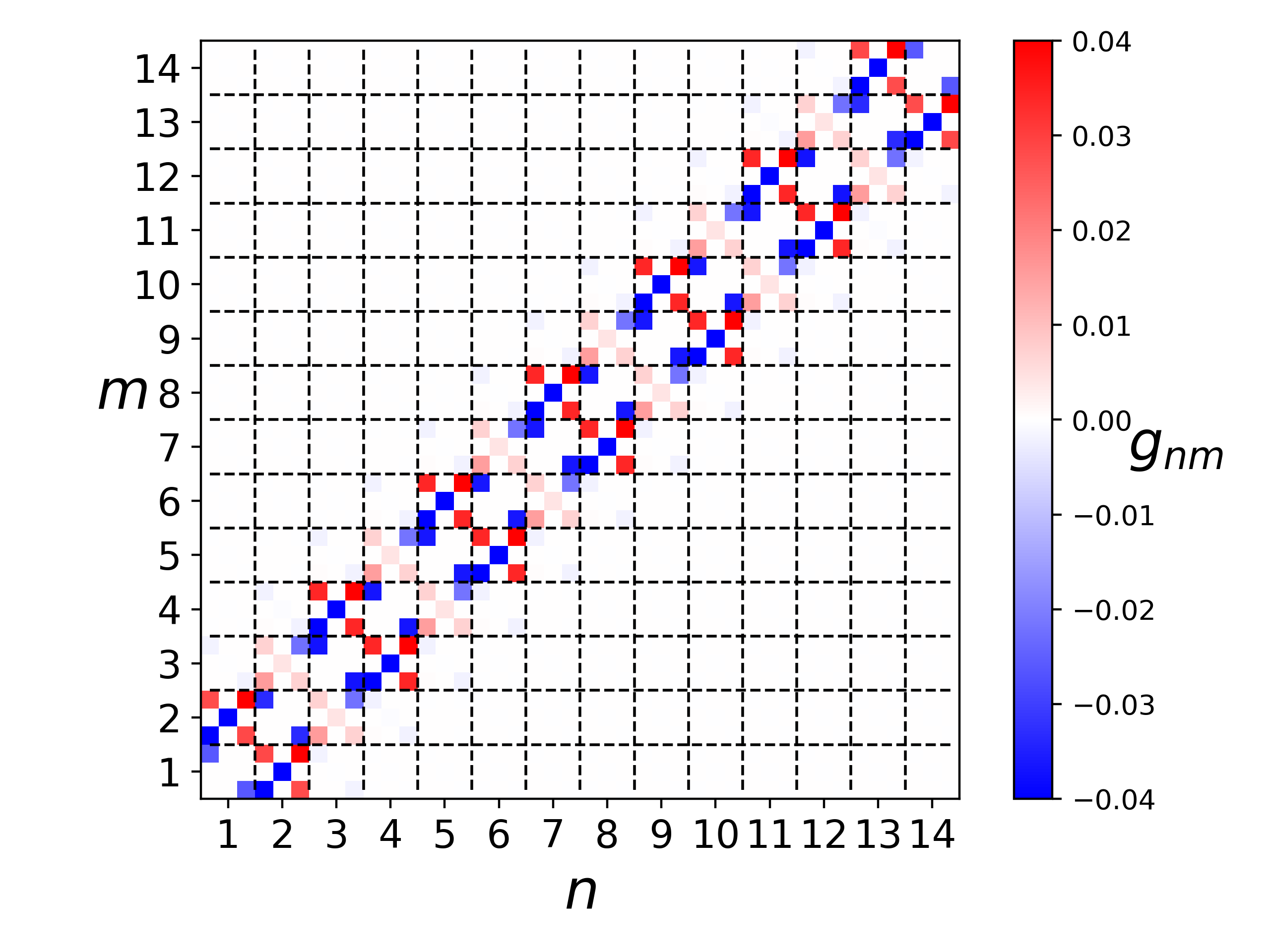}
    \includegraphics[width=0.5\linewidth]{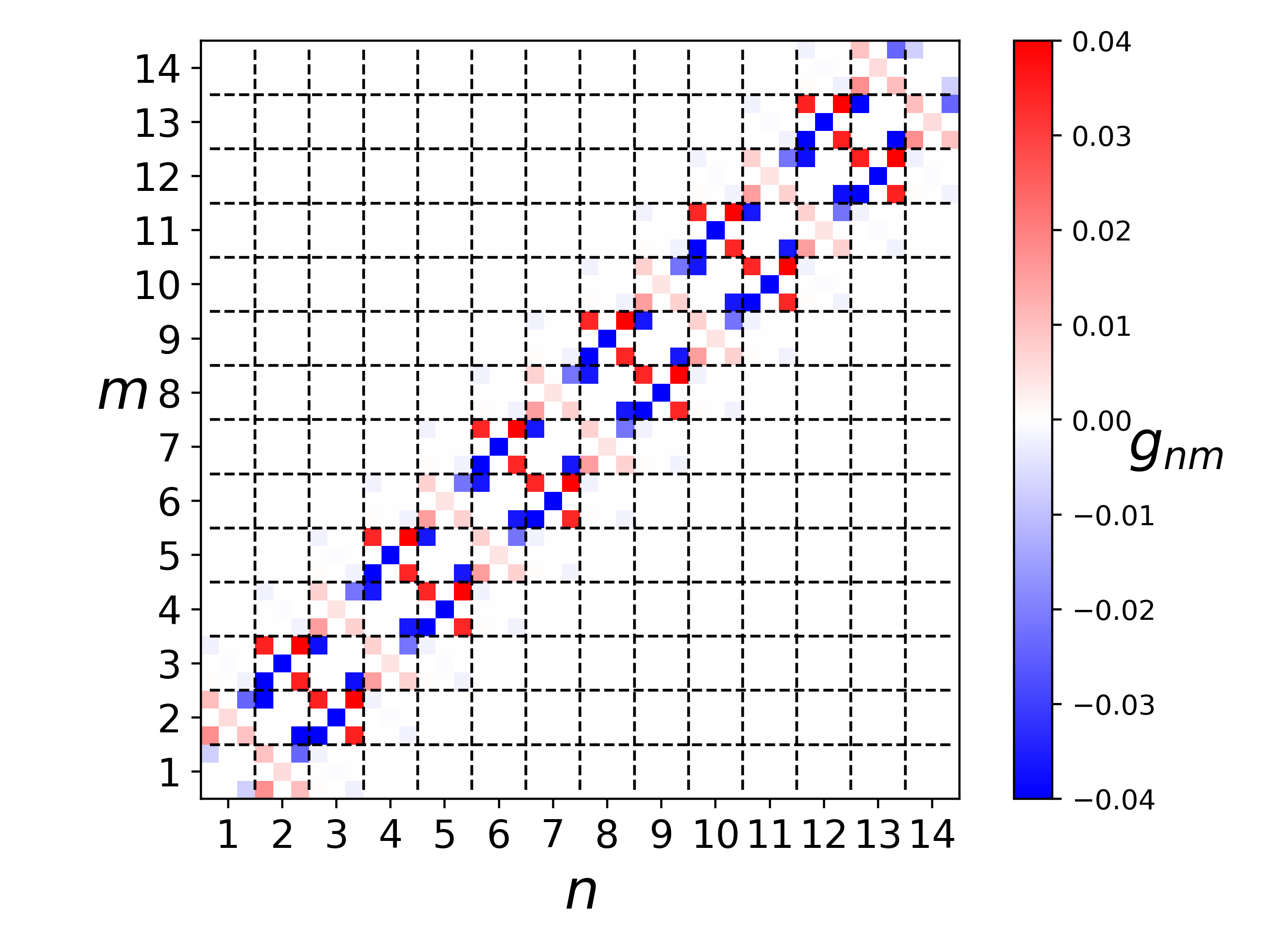}
    \caption{Elements of the interaction matrix $g$ for trivial (top) and topological (bottom) configurations. We saturated the colors to $\pm 0.04$ for a better visualization. Here we fixed $d=2$.}
    \label{fig:5}
\end{figure}

The interaction strength is encoded in the matrix $g$. For each pair $\{n,m\}$ there are nine elements describing the interaction between all directions. We show in Fig.~\ref{fig:5} the elements of the interaction matrix $g$ in a grid. Each box shows nine interactions between the unit cells. In this figure, it is easy to see the staggered nature of the interactions between cells. The main difference between topological and trivial is whether the edges are weakly or strongly interacting with the bulk. 

\section{Atom-phonon coupling}
The details about the analytical derivations of the atom-phonon Hamiltonian can be found in this section. Despite the textbook character, there are important details specific to the problem.

The interaction between the two different species is given by 
\begin{equation}
    H_{\textrm{Ry-a}} = \sum_{n,m}^{N_c}\sum_{\alpha=1}^{N_b} V_{\textrm{Ry-a}}(\vb{r}_n-\vb{R}_{m\alpha}).
\end{equation}
This can be expanded in power series in Rydberg displacements $u$,
\begin{align}
    H_{\textrm{Ry-a}} &= \sum_n\sum_{m,\alpha} V_{\textrm{Ry-a}}(\vb{r} - \vb{R}_{m\alpha}^0 - \vb{u}_{m\alpha}) \nonumber\\
    &\approx \sum_n\sum_{m,\alpha} V_{\textrm{Ry-a}}(\vb{r}_n-\vb{R}_{m\alpha}^0) - \sum_n\sum_{m,\alpha}\vb{u}_{m\alpha}\cdot \grad_{\vb{R}_{m\alpha}}V_{\textrm{Ry-a}}(\vb{r}_n-\vb{R}_{m\alpha})\eval_{\vb{R}_{m\alpha}^0} \nonumber\\
    &\equiv H_{\textrm{Ry-a}}^{(0)} + H_{\textrm{Ry-a}}^{(1)},
\end{align}
with $\vb{R}_{n\alpha}^0 = \vb{R}_n + \vb*{\rho}_{\alpha}$. The constant term in $\vb{u}_{n\alpha}$ forms the periodic potential and it is already taken into account in $\mathcal{H}_{\textrm{a}}$. In that way, the atom-phonon interaction is encoded in the first order term. The constant part will be neglected from here. We will also neglect higher order terms, $\mathcal{O}(u^2)$. The gradient can be straightforwardly evaluated writing $V_{\textrm{Ry-a}}(\vb{r})$ in the reciprocal space as $V_{\textrm{Ry-a}}(\vb{r} - \vb{R}_{n\alpha}) = \frac{1}{V} \sum_{\vb{Q}} e^{i\vb{Q}\cdot(\vb{r} - \vb{R}_{n\alpha})} V_{\vb{Q}}$,
\begin{equation}
   \grad_{\vb{R}_{m\alpha}}V_{\textrm{Ry-a}}(\vb{r}_n-\vb{R}_{m\alpha})\eval_{\vb{R}_{m\alpha}^0}
    = \frac{-i}{V}\sum_{\vb{Q}}\vb{Q} e^{i\vb{Q}\cdot(\vb{r} - \vb{R}_{n\alpha}^0)} V_{\vb{Q}}.
\end{equation}
Thus, we can write the second quantized atom-phonon Hamiltonian \cite{fetter2012quantum},
\begin{equation}
    \mathcal{H}_{\textrm{a-ph}} = \sum_{b,k,\sigma}\sum_{b',k',\sigma'} \bra{b,k,\sigma}H_{\textrm{Ry-a}}^{(1)}\ket{b',k',\sigma'} c_{b,k,\sigma}^{\dagger} c_{b',k',\sigma'},
\end{equation}
where the relevant indexes are the Bloch band $b$, momentum $\vb{k}$, and spin $\sigma$. The coefficients are given by 
\begin{align}\label{eq:braket1}
    \bra{b,k,\sigma}H_{\textrm{Ry-a}}^{(1)}\ket{b',k',\sigma'} &= - \int \dd\vb{r} \Psi^*_{b,k,\sigma}(\vb{r}) \sum_{n,\alpha}\vb{u}_{n\alpha}\cdot \grad_{\vb{R}_{n\alpha}}V_{\textrm{Ry-a}}(\vb{r}-\vb{R}_{n\alpha})\eval_{\vb{R}_{n\alpha}^0} \Psi_{b',k',\sigma'}(\vb{r}) \nonumber\\
    &= \frac{i}{V} \sum_{n,\alpha,\vb{Q}} \vb{Q}\cdot\vb{u}_{n\alpha} e^{-i\vb{Q}\cdot\vb{R}_{n\alpha}^0} V_{\vb{Q}} \int \dd\vb{r} \Psi^*_{b,k,\sigma}(\vb{r}) e^{i\vb{Q}\cdot\vb{r}} \Psi_{b',k',\sigma'}(\vb{r}).
\end{align}
The second quantized displacement operator can be written in terms of its Fourier transform,
\begin{align}
    \vb{u}_{n\alpha} = \sum_{\vb{q}}\sum_{j=1}^{3N_b} \sqrt{\frac{\hslash}{2M_{\alpha}\omega_{j}(\vb{q})}} \vb*{\xi}_{\alpha}^{(j)}(\vb{q}) (b_{j,\vb{q}} + b_{j,\vb{-q}}^{\dagger})\frac{e^{i\vb{q}\cdot\vb{R}_n}}{\sqrt{N}},
\end{align}
where $\omega_{j}$ and $\vb*{\xi}_{\alpha}^{(j)}$ are the eigenvalues and eigenvectors of the dynamical matrix, and $j$ is the phonon band index. 
We can write $e^{-i\vb{Q}\cdot\vb{R}_{n\alpha}^0}=e^{-i\vb{Q}\cdot\vb{R}_{n}}e^{-i\vb{Q}\cdot\vb*{\rho}_{\alpha}}$, and in Eq.~\ref{eq:braket1} will appear $\sum_n e^{i(\vb{q}-\vb{Q})\cdot\vb{R}_{n}} = N\delta_{\vb{q},\vb{Q}}$. 
The atom-phonon Hamiltonian is written as,
\begin{align}
    \mathcal{H}_{\textrm{a-ph}} &= \frac{i}{\sqrt{N}}\sum_{\vb{q},j}\sum_{b,k,\sigma}\sum_{b',k',\sigma'} 
    \sqrt{\frac{\hslash}{2M_{\alpha}\omega_{j}(\vb{q})}} \vb{q}\cdot\vb*{\xi}_{\alpha}^{(j)}(\vb{q}) e^{-i\vb{q}\cdot\vb*{\rho}_{\alpha}} V_{\vb{q}} \alpha_{q,b,k,\sigma,b',k',\sigma'}
    (b_{j,\vb{q}} + b_{j,\vb{-q}}^{\dagger}) c_{b,k,\sigma}^{\dagger} c_{b',k',\sigma'} 
    \nonumber\\
     &= \sum_{\vb{q},j}\sum_{b,k,\sigma}\sum_{b',k',\sigma'} \frac{1}{\sqrt{N}}
    M_{q,j,b,k,\sigma,b',k',\sigma'} (b_{j,\vb{q}} + b_{j,\vb{-q}}^{\dagger}) c_{b,k,\sigma}^{\dagger} c_{b',k',\sigma'},
\end{align}
where we defined 
\begin{equation}
    \alpha_{q,b,k,\sigma,b',k',\sigma'} = \frac{1}{\Omega} \int \dd\vb{r} \Psi^*_{b,k,\sigma}(\vb{r}) e^{i\vb{q}\cdot\vb{r}} \Psi_{b',k',\sigma'}(\vb{r}),
\end{equation}
and $\Omega=V/N$. Let us evaluate the overlap integral above in terms of the Wannier functions within some approximations. First, consider the potential is independent of the spin, so $\sigma=\sigma'$ and then $\alpha$ is independent of the spin. In terms of the Wannier functions,
\begin{equation}
    \Psi_{b,\vb{k}}(\vb{r}) = \frac{1}{\sqrt{N}} \sum_{\vb{R}} e^{i\vb{k}\cdot\vb{R}} \phi_b(\vb{r}-\vb{R}),
\end{equation}
we can evaluate $\alpha$,
\begin{align}
    \Omega\alpha_{q,b,k,b',k'} &= \int \dd\vb{r} \Psi^*_{b,k}(\vb{r}) e^{i\vb{q}\cdot\vb{r}} \Psi_{b',k'}(\vb{r}) \nonumber\\
    &= \frac{1}{N} \sum_{\vb{RR}'} e^{-i\vb{k}\cdot\vb{R}} e^{i\vb{k}'\cdot\vb{R}'}\int \dd\vb{r} e^{i\vb{q}\cdot\vb{r}} \phi^*_b(\vb{r}-\vb{R})\phi_{b'}(\vb{r}-\vb{R}').
\end{align}
Within the tight-binding approximation, terms $\vb{R}\neq\vb{R}'$ are vanishing 
\begin{equation}
    \Omega\alpha_{q,b,k,b',k'} = \frac{1}{N}\sum_{\vb{R}} e^{i(\vb{k}'-\vb{k})\cdot\vb{R}} \int \dd\vb{r} e^{i\vb{q}\cdot\vb{r}} \phi^*_b(\vb{r}-\vb{R})\phi_{b'}(\vb{r}-\vb{R}).
\end{equation}
In the single band approximation, we fix $b=b'=0$,
\begin{align}
    \Omega\alpha_{q,k,k'} &= \frac{1}{N}\sum_{\vb{R}} e^{i(\vb{k}'-\vb{k})\cdot\vb{R}} \int \dd\vb{r} e^{i\vb{q}\cdot\vb{r}} |\phi_0(\vb{r}-\vb{R})|^2
    \nonumber\\
    &= \frac{1}{N}\sum_{\vb{R}} e^{i(\vb{k}'-\vb{k})\cdot\vb{R}} e^{i\vb{q}\cdot\vb{R}} \int \dd\vb{r} e^{i\vb{q}\cdot\vb{r}} |\phi_0(\vb{r})|^2,
\end{align}
but $\frac{1}{N}\sum_{\vb{R}} e^{i(\vb{q}+\vb{k}'-\vb{k})\cdot\vb{R}}=\delta_{\vb{q}+\vb{k}',\vb{k}}$ and define $\rho_0=\int \dd\vb{r} e^{i\vb{q}\cdot\vb{r}} |\phi_0(\vb{r})|^2$. 
The atom-phonon Hamiltonian takes the simple form
\begin{equation}
    \mathcal{H}_{\textrm{a-ph}} = \sum_{\vb{q},\vb{k},j,\sigma} \frac{1}{\sqrt{N}}
    M_{j,\vb{q}} (b_{j,\vb{q}} + b_{j,\vb{-q}}^{\dagger}) c_{\vb{k+q},\sigma}^{\dagger} c_{\vb{k},\sigma},
\end{equation}
where the interaction strength is given by
\begin{equation}
    M_{j,\vb{q}} = \sum_{\alpha} \sqrt{\frac{\hslash}{2M_{\alpha}\omega_{j}(\vb{q})}} \vb{q}\cdot\vb*{\xi}_{\alpha}^{(j)}(\vb{q}) e^{-i\vb{q}\cdot\vb*{\rho}_{\alpha}} V_{\vb{q}} \rho_0(\vb{q}).
\end{equation}
Here we set $\Omega=1$ and get rid of the $i$ by a rotation (global phase transformation).

In a zig-zag chain where $\vb{q}=q\vu{z}$, the dot product in $M_{j\vb{q}}$ restricts the couplings and as a result transverse states without a $z$ component do not interact with the atoms,
\begin{equation}
    \mathcal{H}_{\textrm{a-ph}} = \sum_{q,k,j,\sigma} \frac{1}{\sqrt{N}}
    M_{j,q} (b_{j,q} + b_{j,-q}^{\dagger}) c_{k+q,\sigma}^{\dagger} c_{k,\sigma},
\end{equation}
with 
\begin{equation}
    M_{j q} = \sum_{\alpha} \sqrt{\frac{\hslash g_{cp}^2}{2M_{\alpha}\omega_j(q)}} q\xi_{\alpha,z}^{(j)}(q) e^{-iq\rho_{\alpha}^z} \rho_0(q),
\end{equation}
where a Fermi pseudopotential with magnitude $g_{cp}$ was taken into accound for $V_{q}$. Here we used Ortner \textit{et al.} approximation for the lattice Wannier functions~\cite{Ortner2009}, which for composite lattices may need to be modified~\cite{Marzari1997,Ganczarek2014,Negretti2014} but does not qualitatively impact our results. Thus,
\begin{equation}
    \rho_0(q)=\int \dd \vb{r} e^{i q z}|\phi_{0}(\vb{r})|^2 \approx \frac{8\pi^2 \sin(q d/2)}{4\pi^2 q d - q^3 d^3},
\end{equation}
Therefore, we aim to study the interactions described by the following equation,
\begin{equation}\label{eq:dimM}
    \sqrt{\frac{2M}{\hslash g_{cp}^2}} M_{j q} =\frac{q \rho_0(q)}{\sqrt{\omega_j(q)}}  \sum_{\alpha}|\xi_{\alpha,z}^{(j)}(q)|e^{-iq\rho_{\alpha}^z},
\end{equation}
where $M_{\alpha}=M$. The interaction strength inversely depends on the eigenvalues $\omega_j$ and directly depends on the $z$ component of the eigenvectors $\xi^{(j)}_{\alpha,z}$, while $\rho_0$ and $e^{-iq\rho_{\alpha}^z}$ will just shape the interaction pattern. In other words, to understand the interaction behaviour, one needs both $\omega_j$ and $\xi^{(j)}_{\alpha,z}$.
It is important to note that the eigenvectors obey the following orthonormality condition,
\begin{align}
    \vb*{\xi}^{(j)}\cdot\vb*{\xi}^{(j')} = 
    \xi^{(j)}_{A,x}\xi^{(j')}_{A,x} + \xi^{(j)}_{A,y}\xi^{(j')}_{A,y} + \xi^{(j)}_{A,z}\xi^{(j')}_{A,z} + \xi^{(j)}_{B,x}\xi^{(j')}_{B,x} + \xi^{(j)}_{B,y}\xi^{(j')}_{B,y} + \xi^{(j)}_{B,z}\xi^{(j')}_{B,z}
    = \sum_{\alpha} \vb*{\xi}^{(j)}_{\alpha}\cdot\vb*{\xi}^{(j')}_{\alpha} = \delta_{j,j'}.
\end{align}
If we change $\xi^{(j)}_{\alpha,z} \rightarrow - \xi^{(j)}_{\alpha,z}$ (for all $j$), the orthonormality condition above still holds. That is why there is a modulus in Eq. \ref{eq:dimM}.

\bibliography{refs.bib}{}
\bibliographystyle{unsrt}